\documentclass[aps,twocolumn,preprintnumbers,floats]{revtex4}\def\@cite#1#2{\textsuperscript{[{#1\if@tempswa , #2\fi}]}}

\usepackage{graphicx}
\usepackage{amsmath}
\usepackage{amsfonts}
\usepackage{amssymb}
\usepackage{color}
\usepackage{subfigure}
\usepackage{epsfig}
\usepackage{morefloats}
\usepackage{multirow}
\usepackage{graphicx,booktabs}
\usepackage{mathrsfs}
\usepackage{txfonts}
\usepackage{indentfirst}
\usepackage{graphicx,booktabs}

\usepackage{longtable,lscape}
\usepackage[colorlinks, citecolor=blue,anchorcolor=red,menucolor=red, linkcolor=red,filecolor=red,urlcolor=blue,frenchlinks=red]{hyperref}

\newcommand{\vrho}{\mbox{\boldmath$\rho$\unboldmath}}
\newcommand{\vlab}{\mbox{\boldmath$\lambda$\unboldmath}}

\newcommand{\ot}{\cdot\cdot\cdot}

\begin{document}

\title{$\Omega_c$ baryon spectrum and strong decays in a constituent quark model}

\author{Hui-Hua Zhong$^{1,2}$, Ming-Sheng Liu$^{1}$~\footnote{E-mail: liumingsheng@email.tjut.edu.cn},
Li-Ye Xiao$^{4}$, Kai-Lei Wang$^{5}$, Qi-Li$^{6}$, Xian-Hui Zhong$^{2,3}$~\footnote{E-mail: zhongxh@hunnu.edu.cn}}
\affiliation{ 1) Tianjin Key Laboratory of Quantum Optics and Intelligent Photonics, School of Science, Tianjin University of Technology, Tianjin 300384, China}

\affiliation{ 2) Department of Physics, Hunan Normal University, and Key Laboratory of Low-Dimensional Quantum Structures and Quantum Control of Ministry of Education, Changsha 410081, China }

\affiliation{ 3) Synergetic Innovation Center for Quantum Effects and Applications (SICQEA), Hunan Normal University, Changsha 410081, China}

\affiliation{ 4) Institute of Theoretical Physics, University of Science and Technology Beijing, Beijing 100083, China}

\affiliation{ 5) Department of Electronic Information and Physics, Changzhi University, Changzhi, Shanxi 046011, China}

\affiliation{ 6) School of Science, Tianjin Chengjian University, Tianjin 300000, China}

\begin{abstract}

In this work, we study the $\Omega_c$ baryon spectrum up to the $2P$ excitations
within a semi-relativistic constituent quark potential model, where the mixing between different
configurations with the same spin-parity numbers is dynamically considered.
Furthermore, the strong decay properties for the excited $\Omega_c$ states are evaluated
within an improved chiral quark model by including the relativistic correction term. In a unified framework, we provide
a reasonable explanation of the widths, masses, and mass splittings, for the newly observed $\Omega_c$ resonances
$\Omega_c(3000)$, $\Omega_c(3050)$, $\Omega_c(3065)$, $\Omega_c(3090)$, $\Omega_c(3120)$, $\Omega_c(3185)$, and $\Omega_c(3327)$.
It is found that the configuration mixing is crucial for understanding the strong decay properties and mass splittings,
while the relativistic correction term of the strong transition operator plays an important role in the states dominated by the radial excitations. We expect our study can provide useful references for establishing a more abundant $\Omega_c$ spectrum.

\end{abstract}


\maketitle

\section{Introduction}{\label{introduction}}


For the $\Omega_c$ baryon family, the ground state $\Omega_c(2695)1/2^+$ was first
established in 1985 by the WA62 experiment~\cite{Biagi:1984mu}.
In 2009, its spin partner $\Omega_c(2770)3/2^+$ was also reconstructed
in the $\Omega_c(2695)\gamma$ final state by the Belle collaboration~\cite{Solovieva:2008fw}.
In 2017, five new narrow excited $\Omega_c$ states,
$\Omega_c(3000)$, $\Omega_c(3050)$, $\Omega_c(3065)$, $\Omega_c(3090)$, and $\Omega_c(3120)$
were observed in the $\Xi_c^+K^-$ channel by the LHCb Collaboration~\cite{LHCb:2017uwr}.
Subsequently, except for $\Omega_c(3120)$, the other four states were confirmed by the Belle collaboration~\cite{Belle:2017ext}.
In 2021, the LHCb collaboration observed the $\Omega_c(3000)$, $\Omega_c(3050)$, $\Omega_c(3065)$ and $\Omega_c(3090)$
once again in the process $\Omega_b^-\to \Xi_c^+K^-\pi^- $~\cite{LHCb:2021ptx}. Moreover, they probed the spin
of the four excited $\Omega_c$ states, and found that their spin assignments are consistent with
$J=1/2,3/2,3/2,$ and $5/2$. More recently, besides the confirmation of the five excited $\Omega_c$ states
first observed in 2017, the LHCb Collaboration also observed two new excited states,
$\Omega_c(3185)$ and $\Omega_c(3327)$, in the $\Xi_c^+K^-$ channel~\cite{LHCb:2023sxp}.
The discovery of these new states not only has significantly expanded the $\Omega_c$ family,
but also offers an excellent opportunity to further investigate the strong interactions between heavy and light quarks.

Stimulated by newly observed excited $\Omega_c$ states, many theoretical studies on the
mass spectrum and decay properties have been carried out within various
models and methods~\cite{Ortiz-Pacheco:2020hmj,Garcia-Tecocoatzi:2022zrf,Yu:2023bxn,Luo:2023sra,
Luo:2023sne,Ortiz-Pacheco:2023kjn,Peng:2024pyl,Ortiz-Pacheco:2024qcf,Garcia-Tecocoatzi:2024aqz,
Karliner:2023okv,Yu:2022ymb,Li:2024zze,Weng:2024roa,Patel:2023wbs,Bahtiyar:2020uuj,Padmanath:2017lng,Yang:2021lce,Wang:2023wii,
Pan:2023hwt,Oudichhya:2023awb,Jakhad:2023mni,Jakhad:2024wpx,Jakhad:2024fgt,
Wang:2017vnc,Cheng:2017ove,Huang:2017dwn,Zhao:2017fov,Chen:2017gnu,Luo:2021dvj,Galkin:2020iat,Karliner:2017kfm,Agaev:2017lip,Chen:2017sci,Chen:2015kpa,Wang:2017zjw}.
These newly discovered states are commonly explained as the orbital or radial excitations of the $\Omega_c$ baryon,
however, there remain many controversies for their spin-parity quantum numbers~\cite{Cheng:2021qpd,Li:2024zze}.
On the other hand, the hyperfine mass splitting of the newly observed $\Omega_c(X)$ resonances at LHCb in a very narrow energy
range of $3.0-3.1$ cannot be well described based on various quark potential models.
It should be mentioned that except for conventional three-quark states,
these newly discovered states were also suggested as compact pentaquarks or baryon-meson molecular states in the literature
~\cite{Zhu:2022fyb,Feng:2023ixl,Xin:2023gkf,Yan:2023tvl,Ozdem:2023okg,Ikeno:2023uzz,Huang:2024iua,Ozdem:2024pyb,
Chen:2017xat,Kim:2017jpx,An:2017lwg,Ali:2017wsf,Montana:2017kjw,Debastiani:2017ewu,Santopinto:2018ljf}.
More information about the status of the $\Omega_c$ baryons can be seen in the recent review works~\cite{Cheng:2021qpd,Chen:2022asf}.

In 2017, inspired by the discovery of five new narrow states of $\Omega_c$ at LHCb~\cite{LHCb:2017uwr},
our group investigated the strong and radiative decay properties of the low-lying $\lambda$-mode
orbital ($1P_{\lambda}$) and radial ($2S_{\lambda}$) excitations of $\Omega_c$ within the chiral quark model~\cite{Wang:2017hej}.
In which, the $\Omega_c(3000)$, $\Omega_c(3050)$, $\Omega_c(3065)$, and $\Omega_c(3090)$
are assigned as the $1P_{\lambda}$ states with quantum numbers
$1/2^-$, $3/2^-$, $3/2^-$, and $5/2^-$, respectively;
while the $\Omega_c(3120)$ is assigned as the $2S_{\lambda}$ states with quantum numbers $1/2^+$ or $3/2^+$.
In 2018, our research group extended the studies to the $\lambda$-mode
second orbital ($1D_{\lambda}$) excitations of $\Omega_c$~\cite{Yao:2018jmc}. It is found that
the $1D_{\lambda}$ states have a relatively narrow decay width of a several MeV or several tens MeV,
and the kaonic decay channels $\Xi_cK$, $\Xi_c'K$ and $\Xi_c'^*K$ may be ideal channels for future searches~\cite{Yao:2018jmc}.

In this work, we perform a unified study of both
the mass spectrum and strong decays of the $\Omega_c$ baryons up to the $2P$ states (with one orbital
and one radial excitations at the same time). The main aims of this work are (i)
to provide a comprehensive study of the mass spectrum within a semi-relativistic quark potential model by considering the
dynamical mixing between different configurations including both the $\lambda$- and
$\rho$-mode excitations, which is often neglected in the literature.
Based on the reliable calculations, we expect to understand the hyperfine mass splitting of the
newly observed $\Omega_c(X)$ resonances. (ii) To provide a more reliable predictions of
the strong decays of the excited $\Omega_c$ states within an improved chiral quark model.
For the initial $\Omega_c$ states, we will adopt the genuine wave functions obtained from the potential
model rather than a simple harmonic oscillator (SHO) form as adopted
in the previous works of our group~\cite{Wang:2017hej,Yao:2018jmc}. Furthermore,
in the strong transition operators we will incorporate the relativistic correction term $\mathbf{{\cal H}}^{RC}$,
which was overlooked in the previous studies~\cite{Wang:2017hej,Yao:2018jmc}.
Combining the reliable predictions of the decay properties and mass spectrum,
we expect to give a reasonable explanation for the newly observed $\Omega_c(X)$ resonances.

This paper is organized as follows.
The framework is given in Sec.~\ref{Framework},
where we give an introduction to the potential quark model for the mass calculations
and the chiral quark model for the evaluations of the strong decays.
In Sec.~\ref{result and discussion}, the masses and strong decay properties of
the $\Omega_c$ baryons are presented, while the possible assignments of the newly observed
$\Omega_c$ states are discussed based on the obtained results.
Finally, a summary is given in Sec.~\ref{summary}.

\begin{figure}[ht]
\centering \epsfxsize=6.8 cm \epsfbox{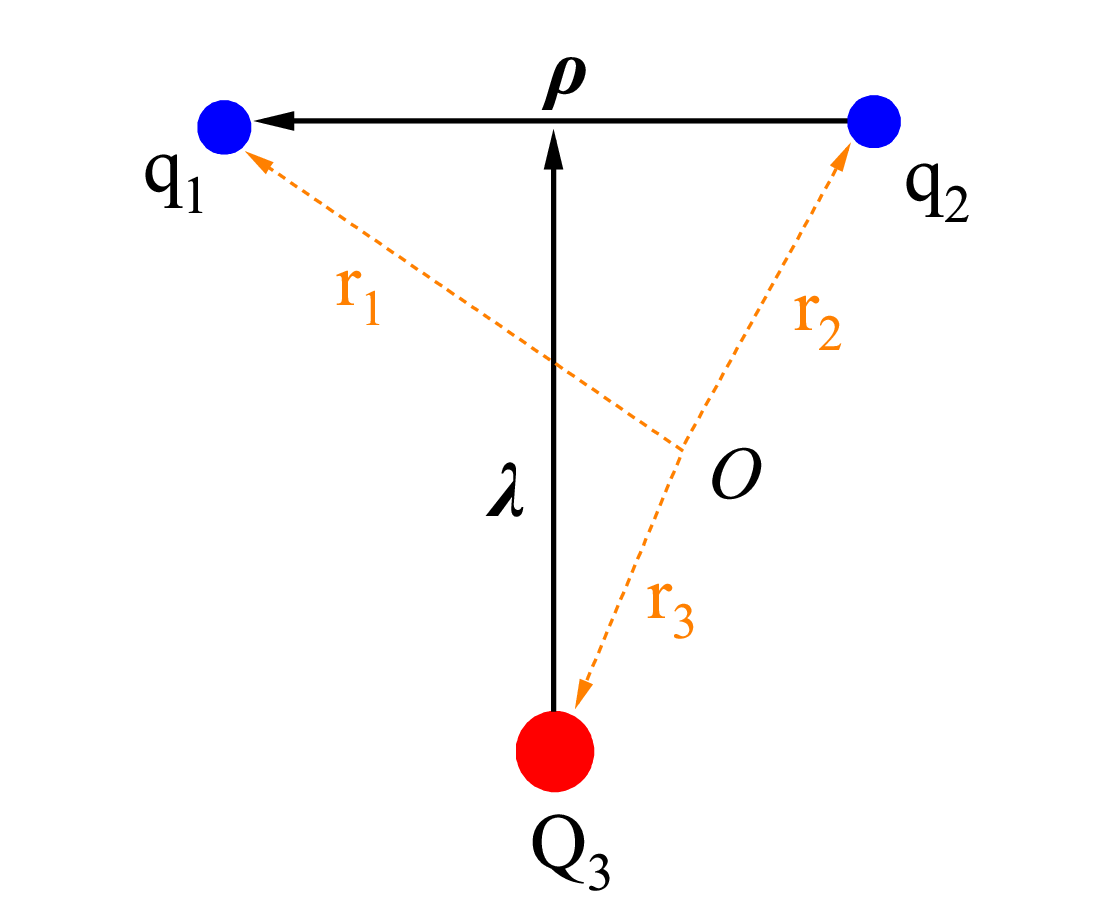}\vspace{-0.2cm} \caption{  Singly heavy baryon system with $\rho$- or $\lambda$-mode excitations.
$q_1$ and $q_2$ stand for the two light quarks, and $Q_3$ stands for the heavy quark.
$\boldsymbol{r}_1, \boldsymbol{r}_2$, and $\boldsymbol{r}_3$ are their position coordinates.}\label{fjcob}
\end{figure}

\section{Framework} \label{Framework}

\subsection{Mass spectrum} \label{mass}

A singly heavy baryon state contains two light quarks and a heavy quark.
The heavy quark violates the flavor SU(4) symmetry. However, the SU(3) symmetry between the other two light quarks is retained.
According to symmetry, the heavy quark is fixed at position $3$, while the two light quarks occupy positions $1$ and $2$.
Their position coordinates are labeled with $\boldsymbol{r}_1, \boldsymbol{r}_2$, and $\boldsymbol{r}_3$ as shown in Fig.~\ref{fjcob}.

\begin{table*}[htp]
\begin{center}
\caption{ The $\Omega_c$ baryon configurations in the $L$-$S$ coupling scheme up to the $2P$ excitation. The configurations are denoted by $n^{2S+1}L_{\lambda(\rho)}{J^P}$, where $n-1$ corresponds to the node number of the wave function.
The right superscripts $(s,\lambda,\rho)$ of the spatial wave function $\psi_{LM_L}$ and spin wave function $\chi$ stand for their permutation symmetries. The Clebsch-Gordan series for the spin and orbital
angular-momentum addition $|JM_J\rangle=\Sigma_{M_{L}+M_{S}=M_J} \langle LM_{L},SM_{S}|J M_J\rangle \ ^n\psi_{LM_L}\chi_{SM_S}$ has been omitted.}\label{TotalWaveFunction}
\setlength{\tabcolsep}{1.5mm}
\begin{tabular}{cccccccccccc}\hline\hline
~~~& Configuration  
~~~& $n$
~~~& $L$
~~~& $S$
~~~& $l_{\rho}$
~~~& $l_{\lambda}$
~~~& $J^{P}$
~~~& Wave function
~~~&                       \\

\hline
~~~& $1^{4}S\frac{3}{2}^{+}$
~~~& $1$
~~~& 0
~~~& $\frac{3}{2}$
~~~& $0$
~~~& $0$
~~~& $\frac{3}{2}^{+}$
~~~& $^1\psi_{00}^{s} \chi^{s}_{\frac{3}{2}M_S} \phi_{f}\phi_{c}$\\

~~~& $1^{2}S\frac{1}{2}^{+}$
~~~& $1$
~~~& 0
~~~& $\frac{1}{2}$
~~~& $0$
~~~& $0$
~~~& $\frac{1}{2}^{+}$
~~~& $^1\psi_{00}^{s} \chi^{\lambda}_{\frac{1}{2}M_S} \phi_{f}\phi_{c}$\\

~~~& $2^{4}S \frac{3}{2}^{+}$
~~~& $2$
~~~& 0
~~~& $\frac{3}{2}$
~~~& $0$
~~~& $0$
~~~& $\frac{3}{2}^{+}$
~~~& $^2\psi_{00}^{s} \chi^{s}_{\frac{3}{2}M_S} \phi_{f}\phi_{c}$\\

~~~& $2^{4}S'\frac{3}{2}^{+}$
~~~& $2$
~~~& 0
~~~& $\frac{3}{2}$
~~~& $0$
~~~& $0$
~~~& $\frac{3}{2}^{+}$
~~~& $^2\psi_{00}^{'s} \chi^{s}_{\frac{3}{2}M_S} \phi_{f}\phi_{c}$\\

~~~& $2^{2}S \frac{1}{2}^{+}$
~~~& $2$
~~~& 0
~~~& $\frac{1}{2}$
~~~& $0$
~~~& $0$
~~~& $\frac{1}{2}^{+}$
~~~& $^2\psi_{00}^{s} \chi^{\lambda}_{\frac{1}{2}M_S} \phi_{f}\phi_{c}$\\

~~~& $2^{2}S' \frac{1}{2}^{+}$
~~~& $2$
~~~& 0
~~~& $\frac{1}{2}$
~~~& $0$
~~~& $0$
~~~& $\frac{1}{2}^{+}$
~~~& $^2\psi_{00}^{'s} \chi^{\lambda}_{\frac{1}{2}M_S} \phi_{f}\phi_{c}$\\

~~~& $1^{4}P_{\lambda }J^{-}$
~~~& $1$
~~~& 1
~~~& $\frac{3}{2}$
~~~& $0$
~~~& $1$
~~~& $\frac{5}{2}^{-},\frac{3}{2}^{-},\frac{1}{2}^{-}$
~~~& $^1\psi_{1M_L}^{\lambda} \chi^{s}_{\frac{3}{2}M_S} \phi_{f}\phi_{c}$\\

~~~& $1^{2}P_{\lambda }J^{-}$
~~~& $1$
~~~& 1
~~~& $\frac{1}{2}$
~~~& $0$
~~~& $1$
~~~& $\frac{3}{2}^{-},\frac{1}{2}^{-}$
~~~& $^1\psi_{1M_L}^{\lambda} \chi^{\lambda}_{\frac{1}{2}M_S} \phi_{f}\phi_{c}$\\

~~~& $1^{2}P_{\rho }J^{-}$
~~~& $1$
~~~& 1
~~~& $\frac{1}{2}$
~~~& $1$
~~~& $0$
~~~& $\frac{3}{2}^{-},\frac{1}{2}^{-}$
~~~& $^1\psi_{1M_L}^{\rho} \chi^{\rho}_{\frac{1}{2}M_S} \phi_{f}\phi_{c}$\\

~~~& $2^{4}P_{\lambda }J^{-}$
~~~& $2$
~~~& 1
~~~& $\frac{3}{2}$
~~~& $0$
~~~& $1$
~~~& $\frac{5}{2}^{-},\frac{3}{2}^{-},\frac{1}{2}^{-}$
~~~& $^2\psi_{1M_L}^{\lambda} \chi^{s}_{\frac{3}{2}M_S} \phi_{f}\phi_{c}$\\

~~~& $2^{4}P'_{\lambda }J^{-}$
~~~& $2$
~~~& 1
~~~& $\frac{3}{2}$
~~~& $0$
~~~& $1$
~~~& $\frac{5}{2}^{-},\frac{3}{2}^{-},\frac{1}{2}^{-}$
~~~& $^2\psi_{1M_L}^{'\lambda} \chi^{s}_{\frac{3}{2}M_S} \phi_{f}\phi_{c}$\\

~~~& $2^{2}P_{\lambda }J^{-}$
~~~& $2$
~~~& 1
~~~& $\frac{1}{2}$
~~~& $0$
~~~& $1$
~~~& $\frac{3}{2}^{-},\frac{1}{2}^{-}$
~~~& $^2\psi_{1M_L}^{\lambda} \chi^{\lambda}_{\frac{1}{2}M_S} \phi_{f}\phi_{c}$\\

~~~& $2^{2}P'_{\lambda }J^{-}$
~~~& $2$
~~~& 1
~~~& $\frac{1}{2}$
~~~& $0$
~~~& $1$
~~~& $\frac{3}{2}^{-},\frac{1}{2}^{-}$
~~~& $^2\psi_{1M_L}^{'\lambda} \chi^{\lambda}_{\frac{1}{2}M_S} \phi_{f}\phi_{c}$\\

~~~& $2^{2}P_{\rho }J^{-}$
~~~& $2$
~~~& 1
~~~& $\frac{1}{2}$
~~~& $1$
~~~& $0$
~~~& $\frac{3}{2}^{-},\frac{1}{2}^{-}$
~~~& $^2\psi_{1M_L}^{\rho} \chi^{\rho}_{\frac{1}{2}M_S} \phi_{f}\phi_{c}$\\

~~~& $2^{2}P'_{\rho }J^{-}$
~~~& $2$
~~~& 1
~~~& $\frac{1}{2}$
~~~& $1$
~~~& $0$
~~~& $\frac{3}{2}^{-},\frac{1}{2}^{-}$
~~~& $^2\psi_{1M_L}^{'\rho} \chi^{\rho}_{\frac{1}{2}M_S} \phi_{f}\phi_{c}$\\

~~~& $1^{4}D_{\lambda }J^{+}$
~~~& $1$
~~~& 2
~~~& $\frac{3}{2}$
~~~& $0$
~~~& $2$
~~~& $\frac{7}{2}^{+},\frac{5}{2}^{+},\frac{3}{2}^{+},\frac{1}{2}^{+}$
~~~& $^1\psi_{2M_L}^{\lambda} \chi^{s}_{\frac{3}{2}M_S} \phi_{f}\phi_{c}$\\

~~~& $1^{2}D_{\lambda }J^{+}$
~~~& $1$
~~~& 2
~~~& $\frac{1}{2}$
~~~& $0$
~~~& $2$
~~~& $\frac{5}{2}^{+},\frac{3}{2}^{+}$
~~~& $^1\psi_{2M_L}^{\lambda} \chi^{\lambda}_{\frac{1}{2}M_S} \phi_{f}\phi_{c}$\\

~~~& $1^{4}D_{\rho }J^{+}$
~~~& $1$
~~~& 2
~~~& $\frac{3}{2}$
~~~& $2$
~~~& $0$
~~~& $\frac{7}{2}^{+},\frac{5}{2}^{+},\frac{3}{2}^{+},\frac{1}{2}^{+}$
~~~& $^1\psi_{2M_L}^{\rho} \chi^{s}_{\frac{3}{2}M_S} \phi_{f}\phi_{c}$\\
~~~& $1^{2}D_{\rho }J^{+}$
~~~& $1$
~~~& 2
~~~& $\frac{1}{2}$
~~~& $2$
~~~& $0$
~~~& $\frac{5}{2}^{+},\frac{3}{2}^{+}$
~~~& $^1\psi_{2M_L}^{\rho} \chi^{\lambda}_{\frac{1}{2}M_S} \phi_{f}\phi_{c}$\\
\hline
\hline
\end{tabular}
\end{center}
\end{table*}

\subsubsection{Hamiltonian} \label{mass}

To properly include the relativistic effects of the light quarks in a singly heavy baryon system, we adopt a semi-relativistic Hamiltonian~\cite{Zhong:2024mnt}, i.e.,
\begin{eqnarray}\label{Total H}
H=\sum_{i=1}^3\sqrt{\boldsymbol{p}_i^2+m_i^2}+\sum_{i<j}V(r_{ij})+C_0,
\end{eqnarray}
where $\boldsymbol{p}_{i}$ and $m_{i}$ stand for the momentum and mass of the $i$-th quark, respectively.
$V(r_{ij})$ represents the effective potentials between the $\emph{i}$-th and $\emph{j}$-th quarks with a distance
$r_{ij}\equiv|\boldsymbol{r}_i-\boldsymbol{r}_j|$, while $C_0$ is the zero point energy.

The internal quark potentials, $V(r_{ij})$, can be decomposed into the spin-independent and spin-dependent parts,
\begin{equation}\label{effective potential}
V(r_{ij})=V^{corn}(r_{ij})+V^{sd}(r_{ij}).
\end{equation}
For the spin-independent part, $V^{corn}(r_{ij})$, we adopt the widely used Cornell form~\cite{Eichten:1978tg}, i.e.,
\begin{equation}\label{Cornell}
V^{corn}(r_{ij})=\frac{b}{2}r_{ij}-\frac{2\alpha_{ij}}{3}\frac{1}{r_{ij}},
\end{equation}
where the first term is the confinement part, while the second term is
the Coulomb-like potential derived from the one gluon exchange~(OGE) model~\cite{Capstick:1986ter,Godfrey:1985xj}.
The $b$ is the slope parameter of the confinement potentials, while $\alpha_{ij}$ are the strong coupling constants.
The spin-dependent part $V^{sd}(r_{ij})$ is given by
\begin{equation}\label{OGE}
V^{sd}(r_{ij})=V^{SS}(r_{ij})+V^{T}(r_{ij})+V^{LS}(r_{ij}),
\end{equation}
where $V^{SS}(r_{ij})$, $V^{T}(r_{ij})$, and $V^{LS}(r_{ij})$ stand for the spin-spin,
tensor, and spin-orbit potentials, respectively.
In the OGE model, the spin-spin and tensor potentials are given by~\cite{Capstick:1986ter,Godfrey:1985xj}
\begin{equation}\label{SS}
V^{SS}(r_{ij})=\frac{2\alpha_{ij}}{3}\left\{\frac{\pi}{2}\cdot\frac{\sigma_{ij}^3
e^{-\sigma_{ij}^2r_{ij}^3}}{\pi^{3/2}}\cdot\frac{16}{3m_im_j}(\boldsymbol{S}_i\cdot\boldsymbol{S}_j)\right\},
\end{equation}
\begin{equation}\label{Tensor}
V^{T}(r_{ij})=\frac{2\alpha_{ij}}{3}\frac{1}{m_im_jr_{ij}^3}\left\{\frac{3(\boldsymbol{S}_i\cdot
\boldsymbol{r}_{ij})(\boldsymbol{S}_j\cdot\boldsymbol{r}_{ij})}{r_{ij}^2}-\boldsymbol{S}_i\cdot\boldsymbol{S}_j\right\},
\end{equation}
while the spin-orbit part is given by~\cite{Capstick:1986ter}
\begin{equation}\label{LS}
V^{LS}(r_{ij})=V_{ij}^{so(v)}+V_{ij}^{so(s)},
\end{equation}
with a color-magnetic piece
\begin{eqnarray}\label{LS,v}
V_{ij}^{so(v)}&=&\frac{1}{r_{ij}}\frac{dV^{Coul}(r_{ij})}{dr_{ij}}\left(\frac{\boldsymbol{r}_{ij}\times\boldsymbol{p}_i\cdot\boldsymbol{S}_i}{2m_i^2}
-\frac{\textbf{r}_{ij}\times\boldsymbol{p}_j\cdot\boldsymbol{S}_j}{2m_j^2}\right.\nonumber\\
&&\left.-\frac{\textbf{r}_{ij}\times\boldsymbol{p}_i\cdot\boldsymbol{S}_i-\boldsymbol{r}_{ij}\times\boldsymbol{p}_j\cdot\boldsymbol{S}_j}{m_im_j}\right),
\end{eqnarray}
and a Thomas-precession piece
\begin{eqnarray}\label{LS,s}
V_{ij}^{so(s)}=-\frac{1}{r_{ij}}\frac{dV^{Conf}(r_{ij})}{dr_{ij}}\left(\frac{\boldsymbol{r}_{ij}\times\boldsymbol{p}_i\cdot\boldsymbol{S}_i}{2m_i^2}
-\frac{\boldsymbol{r}_{ij}\times\boldsymbol{p}_j\cdot\boldsymbol{S}_j}{2m_j^2}\right).
\end{eqnarray}
In the spin-dependent potentials, the $\boldsymbol{S}_i$ represents the spin operator of the $i$th quark.

\subsubsection{Wave function} \label{wF}

The spatial wave function can be set as $\psi(\boldsymbol{r}_1, \boldsymbol{r}_2, \boldsymbol{r}_3)$.
To remove the influence of center-of-mass motion, in the calculations one needs to express
the single-particle coordinates $(\boldsymbol{r}_1, \boldsymbol{r}_2, \boldsymbol{r}_3)$
with the Jacobi coordinates $(\vrho,\vlab,\boldsymbol{R})$ by the following transformation,
\begin{equation}\label{JT}
\left( \begin{array}{ccc} \vrho \\ \vlab \\ \boldsymbol{R}\end{array}\right)\equiv
\left( \begin{array}{ccc} \sqrt{\frac{1}{2}} &-\sqrt{\frac{1}{2}}  & 0\\ \sqrt{\frac{2}{3}}\frac{m_1}{M_{12}} & \sqrt{\frac{2}{3}}\frac{m_2}{M_{12}} & -\sqrt{\frac{2}{3}}\\\sqrt{3}\frac{m_1}{M_{123}} & \sqrt{3}\frac{m_2}{M_{123}}& \sqrt{3}\frac{m_3}{M_{123}}\end{array}\right)
\left( \begin{array}{ccc} \boldsymbol{r}_1 \\ \boldsymbol{r}_2 \\ \boldsymbol{r}_3 \end{array}\right),
\end{equation}
where $M_{12}=m_1+m_2$ and $M_{123}=m_1+m_2+m_3$.
The masses of Jacobian coordinates are defined as $m_{\rho}\equiv 2m_1m_2/M_{12}$, $m_{\lambda}\equiv 3M_{12}m_3/2M_{123}$, and $M_R\equiv M_{123}/3$, respectively.

The spatial wave function expressed with the single-particle coordinates can be
transformed into the form expressed with the Jacobian coordinates, i.e.,
\begin{eqnarray}
\psi(\boldsymbol{r}_1, \boldsymbol{r}_2, \boldsymbol{r}_3)=e^{i\boldsymbol{P}_R\cdot\boldsymbol{R}}
\psi(\vrho,\vlab),
\end{eqnarray}
where $\boldsymbol{P}_R$ is the momentum of the center of mass. For a state with quantum numbers
of total orbital angular momentum $\boldsymbol{L}$ and its $z$ component, $L$ and $M_L$, the wave function of the
relative motion part, $\psi(\vrho,\vlab)$, can be expressed as~\cite{Mitroy:2013eom,Varga:1997xga,Varga:1995dm}
\begin{eqnarray}\label{rwf}
\psi_{LM_L}(\vrho,\vlab)=R(\rho,\lambda)\theta_{LM_L}(\vrho,\vlab),
\end{eqnarray}
where $R(\rho,\lambda)$ and $\theta_{LM}$ stand for the radial and angular parts of the wave function, respectively.
The angular part can be chosen as a vector-coupled product of solid spherical
harmonics of the Jacobi coordinates
\begin{eqnarray}
\theta_{LM_L}(\vrho,\vlab)=\rho^{l_{\rho}}\lambda^{l_{\lambda}}[Y_{l_{\rho}m_{\rho}}(\hat{\vrho})\otimes Y_{l_{\lambda} m_{\lambda}}(\hat{\vlab})]_{LM_L},
\end{eqnarray}
where $l_{\rho(\lambda)}$ and $m_{\rho(\lambda)}$ are the quantum numbers of the orbital angular momentum
and its $z$ component corresponding to $\rho(\lambda)$-mode excitation, respectively.
On the other hand, the radial part $R(\rho,\lambda)$ should be determined by solving the Schr\"{o}dinger equation.

In the spin-orbital ($L$-$S$) coupling scheme, the total wave function for the state with quantum numbers
of total angular momentum and its $z$ component, $J$ and $M_J$, is given by
\begin{eqnarray}
\varPsi_{JM_J}=\mathcal{A}\left\{R(\rho,\lambda)[\theta_{LM_L}(\vrho,\vlab)\otimes\chi_{SM_S}]_{JM_J}\phi_{f}\phi_{c}\right\},
\end{eqnarray}
where $\mathcal{A}$ represents the operator that imposes the antisymmetry on
the total wave function when the two light quarks are exchanged. $\chi$, $\phi_f$, and $\phi_c$
represent the spin, flavor, and color wave functions, respectively.
$S$ and $M_S$ are the quantum numbers of the total spin angular momentum $\boldsymbol{S}$ and its $z$ component, respectively.
The explicit forms of the spin wave function $\chi^{\sigma}_{SM_S}$ with different permutation
symmetries $\sigma(=s, \rho, \lambda)$ can be found in Refs.~\cite{Xiao:2013xi,Wang:2017kfr}.
For a baryon system, the color wave function should be a color singlet. One can explicitly express it as
\begin{eqnarray}
\phi_{c}=\frac{1}{\sqrt{6}}(rgb-rbg+gbr-grb+brg-bgr).
\end{eqnarray}
For the $\Omega_c$ baryons considered in the present work, the flavor wave function is a very simple form, $\phi_{f}=ssc$.
The flavor wave functions for all of the singly heavy baryons can be found in Ref.~\cite{Wang:2017kfr}.
The total wave functions for the $\Omega_c$ baryons up to the $2P$ excitations have been given in Table~\ref{TotalWaveFunction}.

For a physical state, generally, $L$ and $S$ are not good quantum numbers, while the
total angular momentum $\boldsymbol{J}$ and parity are conserved.
Therefore, one should take into account the mixing between different configurations
with the same spin-parity $J^P$ numbers. It should be pointed out that the single heavy baryons should
respect the requirement of the heavy quark symmetry~\cite{Roberts:2007ni,Yoshida:2015tia,Isgur:1991wq,Yamaguchi:2014era}.
In the heavy quark mass limit~($m_Q\to\infty$), the heavy quark spin $\boldsymbol{S}_Q$ is approximately conserved,
since the interactions between the heavy quark and the two light quarks are strongly suppressed by heavy quark mass~\cite{Capstick:1986ter}.
This indicates that, for a singly heavy baryon, the configuration constructed by the $j$-$j$ coupling scheme, $|\{[(l_\rho l_\lambda)_L s_\rho]_j s_Q\}_J\rangle$, may be often closer to the physical state than
that of the $L$-$S$ scheme, $|\{(l_\rho l_\lambda)_L (s_\rho s_Q)_S\}_J\rangle$.
Where $s_\rho$ and $s_Q$ are the quantum numbers of the spin of the two light quark system $\boldsymbol{S}_\rho$
and the heavy quark spin, respectively, and $j$ is the quantum number of angular momentum $\boldsymbol{j}\equiv\boldsymbol{L}+\boldsymbol{S}_\rho$.
According to the properties of angular momentum couplings,
one can establish a relationship between the states constructed within the $L$-$S$ and $j$-$j$ coupling schemes, i.e.,~\cite{Roberts:2007ni}
\begin{align}\label{jj-LS}
&\left|\left\{ \left[\left(l_{\rho}l_{\lambda}\right)_{L}s_{\rho}\right]_{j}s_{Q}\right\}_{J}\right\rangle =(-1)^{L+s_{\rho}+J+\frac{1}{2}}\sqrt{2j+1}\nonumber \\
&\times\sum_{S=|s_{\rho}-s_{Q}|}^{s_{\rho}+s_{Q}} \sqrt{2S+1}
\times \left\{
\begin{array}{ccc}
L & s_{\rho} & j \\
s_{Q} & J & S
\end{array}
\right\}
\left|\left[\left(l_{\rho}l_{\lambda}\right)_{L}\left(s_{\rho}s_{Q}\right)_{S}\right]_{J}\right\rangle,
\end{align}
where $\left\{\begin{matrix}  L & s_{\rho} & j \\ s_{Q} & J & S \end{matrix}\right\}$ is a $6$-$J$ symbol.
For convenience, the state of the $L$-$S$ scheme is labeled with $|^{2S+1}L J^P\rangle$, while that
of the $j$-$j$ scheme is labeled with $|J^P,j~\rangle$.


\subsubsection{Numerical method} \label{Gaus}

In this work, we adopt the variation principle to solve the Schr\"{o}dinger equation.
To calculate the matrix elements in coordinate space, we adopt the explicitly correlated Gaussian method ~\cite{Mitroy:2013eom,Varga:1997xga,Varga:1995dm}. The radial part of the spatial wave function
is expanded in terms of the correlated Gaussian basis.
In the single-particle coordinate system, such a basis function is given by~\cite{Varga:1997xga,Varga:1995dm}
\begin{eqnarray}\label{correlatedGS}
\psi &=& \exp\left(-\sum_{j>i=1}^{3} \alpha_{ij}r_{ij}^2    \right),
\end{eqnarray}
where $\alpha_{ij}$ are variational parameters. Considering the symmetry of the
two light identical quarks, we should have $\alpha_{13}= \alpha_{23}$, thus there are two independent variational parameters,
$\alpha_{12}\equiv a$ and $\alpha_{13}=\alpha_{23} \equiv d$. It should be mentioned that
the form chosen in Eq.~(\ref{correlatedGS}) is simpler for the choice of the nonlinear parameters~\cite{Varga:1997xga}.
To calculate the kinetic energy, it is convenient to use a set of Jacobi
coordinates $\boldsymbol{x}=(\vrho,\vlab)$, instead of the relative vectors $\boldsymbol{r}_{ij}$.
The basis function of Eq.~(\ref{correlatedGS}) can be rewritten as
\begin{eqnarray}\label{correlatedGS2}
G(\boldsymbol{x},A) =e^{\tilde{\boldsymbol{x}}A\boldsymbol{x}},
\end{eqnarray}
where the matrix $A$ is
\begin{eqnarray}\label{Matr}
A=
\left(\begin{array}{cc}
2a+d & 0\\
0 & 3d
\end{array}\right).
\end{eqnarray}
The zero nondiagonal matrix element indicates that for the spatial wave function of the singly heavy baryon containing two light identical quarks,
the Jacobi coordinates $\vrho$ and $\vlab$ can be decoupled.

By using the Gaussian bases, the trial radial part of the spatial wave function $R(\rho,\lambda)$ in Eq.~(\ref{rwf})
can be expanded as
\begin{eqnarray}\label{abc}
R(\rho,\lambda)\equiv R(\boldsymbol{x})=\sum^\mathcal{N}_{k=1}c_k G(\boldsymbol{x},A_k).
\end{eqnarray}
The accuracy of the trial function depends on the length of
the expansion $\mathcal{N}$ and the nonlinear parameters $A_k$. In our
calculations, following the method in Ref.~\cite{Hiyama:2003cu}, we let the
variational parameters form a geometric progression.
The variational parameters $a$ and $d$ are set as
\begin{equation}
\left\{
\begin{array}{l}
a_{\ell}~=a_1~q^{~\ell-1}~~~~~(\ell=1, \cdot\cdot\cdot, \ell_{max}), \\
d_{\ell'}=d_1~q'^{\ell'-1}~~~~(\ell'=1, \cdot\cdot\cdot, \ell'_{max}),
\end{array}
\right.
\end{equation}
where $q$ and $q'$ are the ratio coefficients.
Therefore, six parameters
$\{a_1, a_{\ell_{max}}, \ell_{max}, d_1, d_{\ell'_{max}}, \ell'_{max}\}$
are required to be determined through the variational method. The length of
the expansion $\mathcal{N}$ is determined by $\mathcal{N}=\ell_{max}\times \ell'_{max}$.

For a given state, one can work out the Hamiltonian matrix elements,
\begin{eqnarray}
	H_{kk'}=\langle \varPsi_{JM_J}(\boldsymbol{x},A_k) | H |\varPsi_{JM_J}(\boldsymbol{x},A_{k'}) \rangle,
\end{eqnarray}
where
$\varPsi_{JM_J}(\boldsymbol{x},A_k)=\mathcal{A}\left\{G(\boldsymbol{x},A_k)[\theta_{L}\otimes\chi_{S}]_{JM_J}\phi_{f}\phi_{c}\right\}$ for the $L$-$S$ scheme, while  $\varPsi_{JM_J}(\boldsymbol{x},A_k)=\mathcal{A}\left\{G(\boldsymbol{x},A_k)[[\theta_{L}\otimes \boldsymbol{S}_\rho]_{jM_j}\otimes \boldsymbol{S}_Q]_{JM_J} \phi_{f}\phi_{c}\right\}$ for the $j$-$j$ scheme. Then, by solving the generalized matrix eigenvalue problem,
\begin{eqnarray}
	\sum_{k'=1}^{\mathcal{N}}(H_{kk'}-EN_{kk'})c_{k'}=0,
\end{eqnarray}
one can obtain the eigenenergy $E$, and the expansion coefficients
$\{c_k\}$. $N_{kk'}$ is an overlap factor defined by $N_{kk'}=\langle \varPsi_{JM_J}(\boldsymbol{x},A_k)|\varPsi_{JM_J}(\boldsymbol{x},A_{k'})\rangle$.
For the $\Omega_c$ baryon states, when we take $\ell_{max}=\ell'_{max}=9$, $a_1=0.1$~fm, $a_9=2$~fm, $d_1=0.2$~fm and $d_9=3$~fm,
we can obtain stable solutions.

\begin{table}[htp]
\begin{center}
\caption{\label{model parameters} Quark potential model parameters.}
\scalebox{1.0}{
\begin{tabular}{ccccccccccccccccccc}\hline\hline
~&$m_s$~(GeV) ~&$m_c$~(GeV) ~&$b$~(GeV$^2$)  ~&$C_0$~(GeV) ~&$\alpha_{ss}$\\
~&0.55        ~&1.45        ~&0.12                   ~&$-0.253$      ~&0.78           \\
\hline
~&$\alpha_{sc}$ ~&$\sigma_{ss}$~(GeV) ~&$\sigma_{sc}$~(GeV) ~&$r_{c(ss)}$~(fm) ~&$r_{c(sc)}$~(fm) \\
~&0.57          ~&0.40                ~&0.68                ~&0.230              ~&0.215\\
\hline\hline
\end{tabular}}
\end{center}
\end{table}

\subsubsection{parameters} \label{Para}

There are eight parameters $m_{s(c)}$, $b$, $C_0$, $\alpha_{ss(sc)}$, $\sigma_{ss(sc)}$,
in the quark potential model. We have listed these parameters in Table~\ref{model parameters}.
To be consistent with the previous work of our group~\cite{Ni:2023lvx},
the constituent masses of the strange and charm quark are fixed with $m_s=0.55$ GeV and $m_c=1.45$ GeV, respectively.
The other parameters are determined by fitting the measured masses of
six states: (i) two early established $1S$ states, $\Omega_c(2695)1/2^+$ and $\Omega_c(2770)3/2^+$~\cite{ParticleDataGroup:2022pth};
(ii) four newly observed states $\Omega_c(3000)$, $\Omega_c(3050)$, $\Omega_c(3065)$, and $\Omega_c(3090)$ at LHCb and Belle~\cite{ParticleDataGroup:2022pth}, which are interpreted as the first orbitally excited states with $J^P=1/2^-$, $3/2^-$, $3/2^-$ and $5/2^-$,
respectively according to our previous studies~\cite{Wang:2017hej,Wang:2017kfr}.
The slope parameter $b=0.12$ GeV$^2$ determined in the present work is comparable to
$b=0.11-0.14$ GeV$^2$ adopted in the previous works of our group~\cite{Zhong:2024mnt,li:2021hss,Li:2020xzs,Liu:2019wdr}.

Additionally, it should be pointed out that one cannot obtain stable solutions for
some orbitally excited states due to the divergent behavior of the $1/r^3_{ij}$ terms
in the spin-orbit and tensor potentials, if they are not treated as perturbative terms.
In order to overcome this problem,
following the method adopted in the previous works~\cite{Zhong:2024mnt,Deng:2016stx,Deng:2016ktl,Li:2019tbn},
we introduce two cutoff parameters $r_{c(ss)}$ and $r_{c(sc)}$ in the calculation.
In a small range $(0, r_c)$, we let $1/r^3=1/r_c^3$ in the spin-orbit and tensor potentials.
It is found that the masses of the $1P_{\lambda}$ states of $\Omega_c$ with $J^P=1/2^-$ and $5/2^-$
show some sensitivities to the cutoff distance between the light strange and heavy charm quarks, $r_{c(sc)}$.
Thus, the cutoff parameter $r_{c(sc)}=0.215$~fm is determined by fitting the
measured masses of the $1P$ states $\Omega_c(3000)$ and $\Omega_c(3090)$ together with their mass splitting.
The other cutoff distance between the two light strange quarks, $r_{c(ss)}$, is chosen as
$r_{c(ss)}=0.23$ fm to be consistent with that for the two light $u$/$d$ quarks determined
in our previous work~\cite{Zhong:2024mnt}. It should be mentioned that only the
two $\rho$-mode $1D$-wave configurations $|J^P=1/2^+,1\rangle_{1D_{\rho}}$ and
$|J^P=3/2^+,1\rangle_{1D_{\rho}}$ are notably sensitive to the cutoff distance $r_{c(ss)}$.

\subsection{Strong decay} \label{Strong decay}

In this work, we adopt the chiral quark model~\cite{Manohar:1983md,Li:1994cy,Li:1995si,Li:1997gd,Zhao:2002id,Zhong:2007gp,Zhong:2008kd} to study the OZI-allowed two-body strong decays of the baryon resonances
by emitting a single pseudoscalar meson.
This model has been successfully applied to describe the strong decays of heavy-light mesons and baryons
~\cite{Zhong:2024mnt,Xiao:2013xi,Wang:2017kfr,Yao:2018jmc,Liu:2012sj,Wang:2018fjm,Wang:2019uaj,Wang:2020gkn,
Xiao:2020gjo,Wang:2017hej,Xiao:2018pwe,Liu:2019wdr,Zhong:2009sk,Zhong:2010vq,Xiao:2014ura,Ni:2021pce,Ni:2023lvx,li:2021hss}.
The effective low energy quark-pseudoscalar-meson coupling in the SU(3) flavor
basis at tree level is given by~\cite{Manohar:1983md}
\begin{eqnarray}\label{effective coupling}
\mathcal{L}_{PS}=\frac{\delta}{\sqrt{2}f_{\mathbb{M}}}\bar{\psi}_j\gamma_{\mu}\gamma_5\psi_j\vec{I}\cdot\partial^\mu\vec{\phi}_{\mathbb{M}}.
\end{eqnarray}
In Eq.~(\ref{effective coupling}), $\psi_j$ corresponds to the $j$th light quark field in a hadron,
$I$ is a flavor operator, $\phi_{\mathbb{M}}$ denotes the pseudoscalar meson octet,
$f_{\mathbb{M}}$ stands for the pseudoscalar meson decay constant,
and $\delta$ is a global parameter accounting for the strength of the quark-pseudoscalar-meson couplings.

To match the nonrelativistic baryon wave functions in the calculations,
it is crucial to adopt a nonrelativistic form of the effective Lagrangian in the calculations.
Performing a nonrelativistic reduction up to the mass order of $1/m^2$, one has~\cite{Ni:2023lvx,Zhong:2024mnt}
\begin{eqnarray}\label{Hi}
H_I=\mathcal{H}^{NR}+\mathcal{H}^{RC},
\end{eqnarray}
with
\begin{eqnarray}\label{Hnr}
\mathbf{{\cal H}}^{NR}=g\sum_{j}\left(\mathcal{G}\boldsymbol{\sigma}_{j}\cdot\boldsymbol{q}
+\frac{\omega_{\mathbb{M}}}{2\mu_{q}}\boldsymbol{\sigma}_{j}\cdot\boldsymbol{p}_{j}\right )F(\boldsymbol{q}^2)I_{j}\varphi_{\mathbb{M}},
\end{eqnarray}
and
\begin{eqnarray}\label{Hrc}
\mathbf{{\cal H}}^{RC}  &=&-\frac{g}{32\mu_{q}^{2}}\underset{j}{\sum}\left[m_{\mathbb{P}}^2(\boldsymbol{\sigma}_{j}\cdot\boldsymbol{q})\right.\nonumber\\
&&\left.+2\boldsymbol{\sigma}_{j}\cdot(\boldsymbol{q}-2\boldsymbol{p}_{j})\times(\boldsymbol{q}\times\boldsymbol{p}_{j})\right]
F(\boldsymbol{q}^2)I_{j}\varphi_{\mathbb{M}}.
\end{eqnarray}
Here, $\mathbf{{\cal H}}^{NR}$ is a nonrelativistic term up to the $1/m$ order, while $\mathbf{{\cal H}}^{RC}$
is a relativistic correction term at the mass order of $1/m^2$.
In the above equations, $\boldsymbol{\sigma}_{j}$ and $\boldsymbol{p}_{j}$ are the spin
operator and internal momentum operator of the $j$-th light quark within a hadron.
$\varphi_{\mathbb{M}}=e^{-i\boldsymbol{q}\cdot\boldsymbol{r}_j}$ is the plane wave part of the emitted light meson.
The factors $g$ and $\mathcal{G}$ are defined by $g=\delta\sqrt{(E_i+M_i)(E_f+M_f)}/(\sqrt{2}f_{\mathbb{M}})$
and $\mathcal{G}=-\left(\frac{\omega_{\mathbb{M}}}{E_f+M_f}+1+\frac{\omega_{\mathbb{M}}}{2m'_j}\right)$,
in which $(E_i, M_i)$ and $(E_f, M_f)$ are the energy and mass of the initial baryon and final baryon, respectively.
$\omega_{\mathbb{M}}$, $\boldsymbol{q}$, and $m_{\mathbb{P}}$ are the energy,
three momentum and mass of the final state pseudoscalar meson.
The reduced mass $\mu_q$ is expressed as $1/\mu_q=1/m_j+1/m'_j$ for the masses
of the $j$-th quark in the initial and final baryons, respectively.
The factor $F(\boldsymbol{q}^2)=\sqrt{\frac{\Lambda^2}{\Lambda^2+\boldsymbol{q}^2}}$
is introduced to suppress the unphysical contributions in the high momentum region,
where $\Lambda$ is taken of the order of the chiral symmetry breaking scale.
The flavor operators $I_j$ for each pseudoscalar meson together with their matrix representations are given in Table~\ref{Flavor Operators}.

\begin{table*}[htp]
\begin{center}
\caption{\label{Flavor Operators} Flavor operators $I_j$ and their Matrix representations. $a^{\dagger}(u, d, s)$ and $a(u, d, s)$ are the creation and annihilation operators for the $u$, $d$, and $s$ quarks. $\phi_P$ is the mixing angle between the flavor eigenstates $\eta_q(=\frac{1}{\sqrt{2}}(u\bar{u}+d\bar{d}))$ and $\eta_s(=s\bar{s})$. $\lambda^i$ ($i=1-8$) are the Gell-Mann Matrices, and $\lambda^0$ is a $3\times 3$ identity matrix. }
\scalebox{1.0}{
\begin{tabular}{ccccccccccccccccccc}\hline\hline
~~~~~~~~~&meson
~~~~~~~~~&$I_j$
~~~~~~~~~&Matrix representation of $I_j$\\
\hline

~~~~~~~~~&$\pi^+$
~~~~~~~~~&$a_j^{\dagger}(d)a_j(u)$
~~~~~~~~~&$(\lambda^1-i\lambda^2)/2$\\

~~~~~~~~~&$\pi^-$
~~~~~~~~~&$a_j^{\dagger}(u)a_j(d)$
~~~~~~~~~&$(\lambda^1+i\lambda^2)/2$\\

~~~~~~~~~&$\pi^0$
~~~~~~~~~&$\frac{1}{\sqrt{2}}[a_j^{\dagger}(u)a_j(u)-a_j^{\dagger}(d)a_j(d)]$
~~~~~~~~~&$\lambda^3/\sqrt{2}$\\

~~~~~~~~~&$K^+$
~~~~~~~~~&$a_j^{\dagger}(s)a_j(u)$
~~~~~~~~~&$(\lambda^4-i\lambda^5)/2$\\

~~~~~~~~~&$K^0$
~~~~~~~~~&$a_j^{\dagger}(s)a_j(d)$
~~~~~~~~~&$(\lambda^6-i\lambda^7)/2$\\

~~~~~~~~~&$\bar{K}^0$
~~~~~~~~~&$a_j^{\dagger}(d)a_j(s)$
~~~~~~~~~&$(\lambda^6+i\lambda^7)/2$\\

~~~~~~~~~&$K^-$
~~~~~~~~~&$a_j^{\dagger}(u)a_j(s)$
~~~~~~~~~&$(\lambda^4+i\lambda^5)/2$\\

~~~~~~~~~&$\eta$
~~~~~~~~~&$\frac{1}{\sqrt{2}}[a_j^{\dagger}(u)a_j(u)+a_j^{\dagger}(d)a_j(d)]\cos\phi_P-a_j^{\dagger}(s)a_j(s)\sin\phi_P$
~~~~~~~~~&$(\frac{\sqrt{2}}{3}\lambda^0+\frac{1}{\sqrt{6}}\lambda^8)\cos\phi_P-(\frac{1}{3}\lambda^0-\frac{1}{\sqrt{3}}\lambda^8)\sin\phi_P$\\

~~~~~~~~~&$\eta'$
~~~~~~~~~&$\frac{1}{\sqrt{2}}[a_j^{\dagger}(u)a_j(u)+a_j^{\dagger}(d)a_j(d)]\sin\phi_P+a_j^{\dagger}(s)a_j(s)\cos\phi_P$
~~~~~~~~~&$(\frac{\sqrt{2}}{3}\lambda^0+\frac{1}{\sqrt{6}}\lambda^8)\sin\phi_P+(\frac{1}{3}\lambda^0-\frac{1}{\sqrt{3}}\lambda^8)\cos\phi_P$\\

\hline\hline
\end{tabular}}
\end{center}
\end{table*}

It should be noted that the strong decay properties
of the excited $\Omega_c$ states have been studied within the chiral quark model~\cite{Wang:2017kfr,Yao:2018jmc,Wang:2017hej}.
In these works, only the $\mathbf{{\cal H}}^{NR}$ term is kept in the calculations,
while the relativistic correction term $\mathbf{{\cal H}}^{RC}$ is overlooked.
Recently, the $\mathbf{{\cal H}}^{RC}$ term was first included in the investigations
of the strong decays of heavy baryon resonances and multistrangeness baryon resonances~\cite{Arifi:2021orx,Arifi:2022ntc}.
Subsequently, in the study of the decays of heavy-light mesons, non-strange light baryons,
the effects of the $\mathbf{{\cal H}}^{RC}$ term were also taken account for by our group~\cite{Ni:2023lvx,Zhong:2024mnt}.
It is found that the agreement with the data is significantly improved.
In this study, to enhance the reliability of the results, we include the effects of
the $\mathbf{{\cal H}}^{RC}$ term.

By using the transition operator given in Eq.~(\ref{Hi}), the two-body OZI-allowed strong decay amplitude for the $\mathcal{B}\to\mathcal{B}'\mathbb{M}$ process can be worked out by
\begin{eqnarray}\label{amp}
\mathcal{M}_{J_{iz}J_{fz}}[\mathcal{B}\to\mathcal{B}'\mathbb{M}]=\langle\mathcal{B}'_{J_{fz}}|H_I|\mathcal{B}_{J_{iz}}\rangle,
\end{eqnarray}
where $\mathcal{B}$ and $\mathcal{B}'$ stand for the initial and final baryon states, respectively,
and $\mathbb{M}$ is the emitting pseudoscalar meson. $J_{iz}$ and $J_{fz}$  represent the third components of the total
angular momenta of the initial and final baryons, respectively. With derived decay amplitudes from Eq.~(\ref{amp}),
the partial decay width for the $\mathcal{B}\to\mathcal{B}'\mathbb{M}$ process can be obtained with
\begin{eqnarray}\label{width}
\Gamma=\frac{1}{8\pi}\frac{|\boldsymbol{q}|}{M^2_i}\frac{1}{2J_i+1}
\sum_{J_{iz}J_{fz}}|\mathcal{M}_{J_{iz}J_{fz}}|^2,
\end{eqnarray}
where $J_i$ is the total angular momentum quantum number of the initial baryon.


The parameters of the chiral quark model have been well determined in
our previous studies~\cite{Zhong:2024mnt,Zhong:2007gp,Zhong:2008kd,Ni:2023lvx}. The decay constants for $K$ and $\eta$ are
chosen as $f_{K/\eta}=0.113$~MeV. The cut-off parameter $\Lambda$ in the factor
$F(\boldsymbol{q}^2)=\sqrt{\frac{\Lambda^2}{\Lambda^2+\boldsymbol{q}^2}}$
is taken as $\Lambda=0.66$~GeV according to our recent studies in Refs.~\cite{Zhong:2024mnt,Ni:2023lvx}.
The overall dimensionless parameter $\delta$ accounting for the strength of the quark-meson couplings
is fixed with $\delta=0.576$. The mixing angle between the flavor eigenstates $\eta_q$ and $\eta_s$
is taken as $\phi_P=41.2^\circ$ as that determined in Ref.~\cite{Zhong:2011ht}.
Additionally, in this work, the constituent quark mass for the $u/d$ quark is taken with $m_{u/d}=0.42$~GeV,
within the typical range of $0.4\pm0.1$~GeV.

In the calculations of the strong decays, we also need the
masses and wave functions of initial and final baryon states.
In this work, the masses and wave functions of the excited $\Omega_c$
states are taken from our quark model predictions. It should be mentioned that
the radial wave functions of the initial $\Omega_c$ states directly adopt
the numerical form given in Eq.~(\ref{abc}), which are expanded with Gaussian bases.
Where, the expansion coefficients and the variational parameters have been
determined by the variation principle.
On the other hand, for the final baryon states, $\Xi_c$, $\Xi_c'$, $\Xi_c'^*$, $\Xi_c(2790)$, and $\Xi_c(2815)$,
their radial wave functions are adopted the simple harmonic oscillator form,
\begin{equation}\label{spatialfunction}
\begin{aligned}
R_{n_{\xi}l_{\xi}}(\alpha_{\xi},\xi)
&=\alpha_{\xi }^{\frac{3}{2}}\left[\frac{2^{l_{\xi}+2-n_{\xi}}\left(2l_{\xi}+2n_{\xi}+1\right)!!}{\sqrt{\pi}n_{\xi}!\left[(2l_{\xi}+1)!!\right]^{2}}\right]^{\frac{1}{2}}\\
&\times(\alpha_{\xi }\xi)^{l}e^{-\frac{1}{2}\alpha_{\xi }^{2}\xi^{2}}F\left(-n_{\xi},l_{\xi}+\frac{3}{2},\alpha_{\xi }^{2}\xi^{2}\right),
\end{aligned}
\end{equation}
where $F\left(-n_{\xi},l_{\xi}+\frac{3}{2},\alpha_{\xi }^{2}\xi^{2}\right)$ ($\xi=\rho,\lambda$) is the confluent hypergeometric function.
The harmonic oscillator parameter $\alpha_\rho$ for the $\rho$ mode
in the radial wave function is taken as $\alpha_\rho=0.49$~GeV,
while the parameter $\alpha_\lambda$ for the $\lambda$ mode is
related to $\alpha_\rho$ by
$\alpha_\lambda=\left(\frac{3m_c}{m_u+m_s+m_c}\frac{(m_u+m_s)^2}{4m_um_s}\right)^{1/4}\alpha_\rho$~\cite{Zhong:2007gp}.
The masses of the meson and baryon states in the final states
are taken from the PDG~\cite{ParticleDataGroup:2022pth}.
For clarity, the parameters adopted in the present work have been collected in Table~\ref{Chiralp}.

\begin{table}[htp]
\begin{center}
\caption{\label{Chiralp} Parameters for strong decays.}
\scalebox{1.0}{
\begin{tabular}{ccccccccccccccccccc}\hline
\hline
\multicolumn{6}{c}{Chiral quark model parameters}\\
\hline
&$f_{K/\eta}$~(GeV) ~&$\Lambda$~(GeV) ~&$\delta$  ~&$\phi_P$      ~&$m_{u/d}$~(GeV) \\
&0.113             ~&0.66            ~&0.576     ~&$41.2^\circ$  ~&0.42 \\
\hline
\multicolumn{6}{c}{Masses (MeV) for the final states }\\
\hline
&$\Xi_c$
~~~~&$\Xi_c'$
~~~~&$\Xi_c'^*$
~~~~&$\Xi_c(2790)$
~~~~&$\Xi_c(2815)$\\


&2468
~~~~&2578
~~~~&2645
~~~~&2792
~~~~&2817\\

\hline
&$\Omega_c$&$\Omega_c(2770)$&$K$
~~~~&$\eta$\\
& 2695
& 2766
&494
~~~~&548\\
\hline\hline
\end{tabular}}
\end{center}
\end{table}

\begin{figure*}[htbp]
 \centering \epsfxsize=12.6 cm \epsfbox{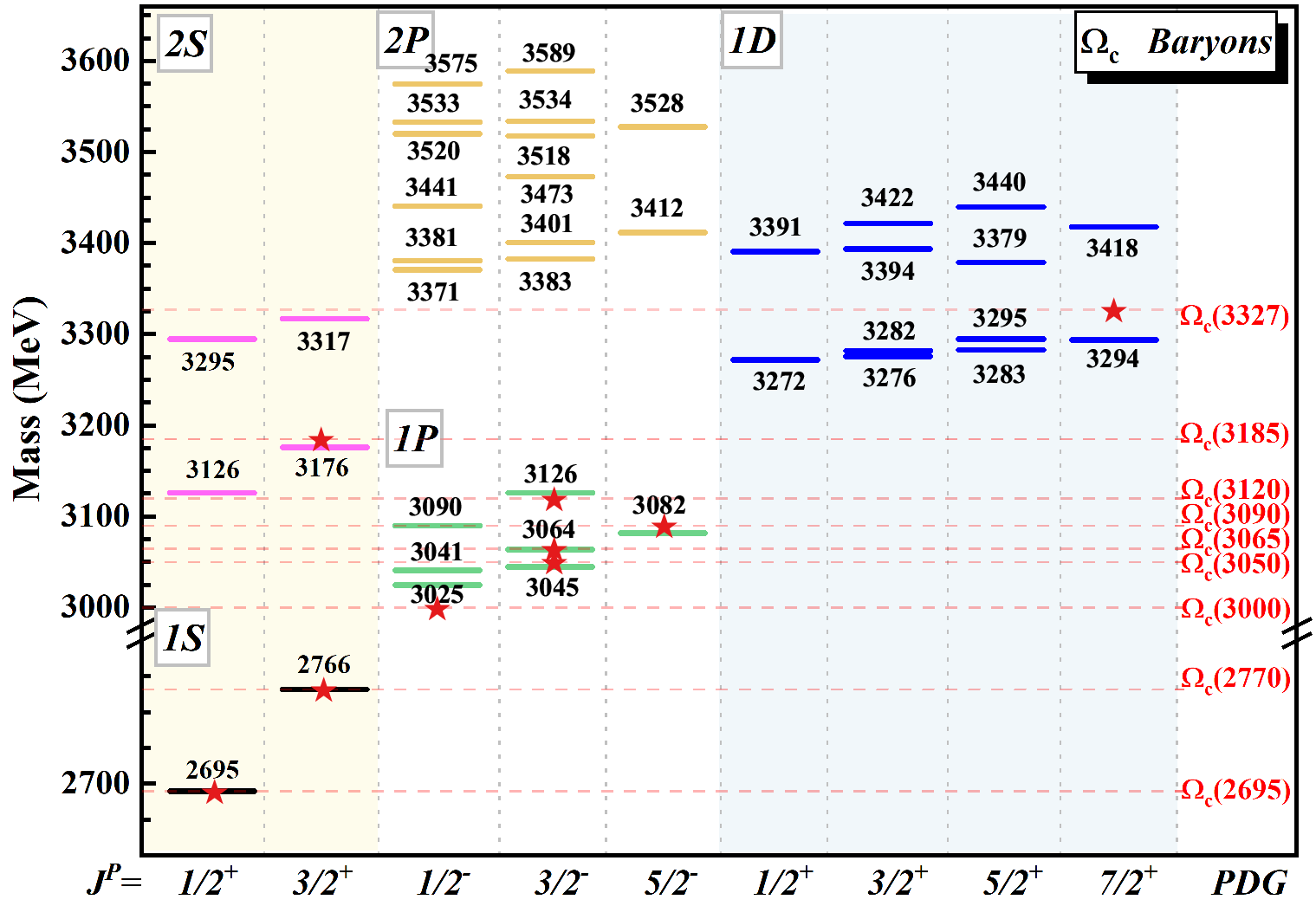}
 \caption{$\Omega_c$ baryon spectrum. It should be mentioned that the $2S$ states with masses of 3295 and 3317 MeV have sizeable
 $1D$ components, the details can be seen in Table~\ref{Omegac mass spectrum comparison}.}\label{omegac mass spectrum}
\end{figure*}

\section{results and discussions} \label{result and discussion}

The obtained mass spectrum is given in Table~\ref{Omegac mass spectrum comparison},
and also plotted in Fig~\ref{omegac mass spectrum}. It should be mentioned that in the mass spectrum calculations,
we consider the dynamical mixing between different configurations of the $j$-$j$ coupling scheme with the same spin-parity numbers.
In Table~\ref{Omegac mass spectrum comparison}, the eigenvectors are given if there is significant configuration mixing.
Furthermore, by combining the spectrum obtained within the potential model,
the strong decay properties of the excited $\Omega_c$ baryons are studied within the chiral quark model.
The decay properties are given in Table~\ref{Omegac decay widths}.

\begin{table*}
\begin{center}
\caption{ The mass spectrum of $\Omega_c$ baryons compared with the observations. The
mixing between different configurations with the same spin-parity numbers are dynamically considered,
and the eigenvectors are given if there is significant configuration mixing.  }
\label{Omegac mass spectrum comparison}
\begin{tabular}{cccccccccccccccccccccccccc}\hline\hline
\tabcolsep=0.5cm
 ~~~~~Configuration~~~~~
& ~~~~~$\langle H \rangle$ (MeV)~~~~~
& ~~~~~Eigenvector~~~~~
& ~~~~~Mass (MeV)~~~~~
&~~~~~Observed state~~~~~

\\
\hline
$|J^P=1/2^+,1\rangle_{1S }$   &(2695)  &1 &2695 (input)  &$\Omega_c(2695)$\\
$|J^P=3/2^+,1\rangle_{1S }$   &(2766)  &1 &2766 (input) &$\Omega_c(2766)$\\
\hline
$|J^P=1/2^-,1\rangle_{1P_\lambda}$
                                &\multirow{3}{*}{$\left(\begin{array}{cccccc}   3028 &-4 &-11  \\   -4 & 3040 & 2\\ -11 & 2& 3088\\ \end{array}\right)$}
                                &\multirow{3}{*}{$\left(\begin{array}{cccccc}   0.96 &0.24  &0.15 \\   -0.22 &0.97 &-0.09\\ -0.17 &0.05  &0.98 \\ \end{array}\right)$}
                                &\multirow{3}{*}{$\left(\begin{array}{cccccc}   3025~ (\mathrm{input})   \\   3041 \\  3090\\ \end{array}\right)$}
                                &$\Omega_c(3000)$\\
$|J^P=1/2^-,0\rangle_{1P_\lambda}$  & &    &\\
$|J^P=1/2^-,1\rangle_{1P_\rho}$     & &    &\\

$|J^P=3/2^-,2\rangle_{1P_\lambda}$
           &\multirow{3}{*}{$\left(\begin{array}{cccccc}    3058 &7 &7 \\ 7 &3065 &-28 \\ 7 &-28 &3113\\ \end{array}\right)$}
           &\multirow{3}{*}{$\left(\begin{array}{cccccc}    0.81 &0.55 &0.21  \\   -0.59 &0.72 &0.36 \\ 0.05 &-0.42 &0.91\\ \end{array}\right)$}
           &\multirow{3}{*}{$\left(\begin{array}{cccccc}    3064~ (\mathrm{input})\\   3045~ (\mathrm{input}) \\ 3126\\ \end{array}\right)$}
           &$\Omega_c(3065)$\\
$|J^P=3/2^-,1\rangle_{1P_\lambda}$  & &    & &$\Omega_c(3050)$\\
$|J^P=3/2^-,1\rangle_{1P_\rho}$     & &    & &$\Omega_c(3120)$\\

 $|J^P=5/2^-,2\rangle_{1P_\lambda}$&(3082)  &1 &3082 (input) &$\Omega_c(3090)$\\
\hline
$|J^P=1/2^+,1\rangle_{2S }$   &(3126)   &1    &3126\\
$|J^P=1/2^+,1\rangle_{2S' }$
          &\multirow{3}{*}{$\left(\begin{array}{cccccc}  3306 &2 &38 \\ 2 &3281 &-21 \\ 38 &-21 &3370\\ \end{array}\right)$}
          &\multirow{3}{*}{$\left(\begin{array}{cccccc}    0.81 &0.52 &-0.26  \\   -0.42 &0.84 &0.34 \\ 0.40 &-0.17 &0.90\\ \end{array}\right)$}
          &\multirow{3}{*}{$\left(\begin{array}{cccccc}    3295 \\   3272 \\ 3391\\\end{array}\right)$}\\
$|J^P=1/2^+,1\rangle_{1D_{\lambda }}$  &  &   &\\
$|J^P=1/2^+,1\rangle_{1D_{\rho }}$  &  &   &\\

$|J^P=3/2^+,1\rangle_{2S }$  & (3176)   &1    &3176 &$\Omega_c(3185)$\\
$|J^P=3/2^+,1\rangle_{2S' }$
          &\multirow{5}{*}{$\left(\begin{array}{cccccc}  3324 &-2 &-1 &-1 &35 \\ -2 &3286 &-6 &-25 &4 \\ -1 &-6 &3300 &4 &-44\\ -1 &-25 &4 &3395 &-11 \\35 &4 &-44 &-11 &3385\\ \end{array}\right)$}
          &\multirow{5}{*}{$\left(\begin{array}{cccccc}    0.88 &-0.11 &0.43 &-0.07 &-0.17 \\
          -0.25 &-0.84 &0.38 &-0.17 &0.26 \\ 0.23 &-0.49 &-0.75 &-0.11 &-0.36\\
          0.20 &-0.17 &-0.15 &0.85 &0.43 \\  0.27 &0.12 &-0.30 &-0.48 &0.77 \\ \end{array}\right)$}
          &\multirow{5}{*}{$\left(\begin{array}{cccccc}    3317 \\   3282\\ 3276\\ 3394\\ 3422\\\end{array}\right)$}\\
$|J^P=3/2^+,2\rangle_{1D_{\lambda }}$   &  &   &\\
$|J^P=3/2^+,1\rangle_{1D_{\lambda }}$   &  &   &\\
$|J^P=3/2^+,2\rangle_{1D_{\rho }}$   &  &   &\\
$|J^P=3/2^+,1\rangle_{1D_{\rho }}$   &  &   &\\

$|J^P=5/2^+,3\rangle_{1D_{\lambda }}$
          &\multirow{4}{*}{$\left(\begin{array}{cccccc}  3293 &2 &-23 &7 \\ 2 &3304 &7 &-40
          \\ -23 &7 &3372 &5 \\ 7 &-40 &5 &3428\\ \end{array}\right)$}
          &\multirow{3}{*}{$\left(\begin{array}{cccccc}    -0.80 &0.52 &-0.26 &0.19 \\
          0.55 &0.81 &0.08 &0.21 \\ 0.26 &-0.08 &-0.96 &-0.01 \\ 0.04 &-0.28 &0.03 &0.96\\ \end{array}\right)$}
          &\multirow{4}{*}{$\left(\begin{array}{cccccc}    3283 \\   3295 \\ 3379 \\ 3440 \end{array}\right)$}\\
$|J^P=5/2^+,2\rangle_{1D_{\lambda }}$  &  &   &\\
$|J^P=5/2^+,3\rangle_{1D_{\rho }}$  &  &   &\\
$|J^P=5/2^+,2\rangle_{1D_{\rho }}$  &  &   &\\

$|J^P=7/2^+,3\rangle_{1D_{\lambda }}$
          &\multirow{2}{*}{$\left(\begin{array}{cccccc}  3300 &-26 \\ -26 &3412\\ \end{array}\right)$}
          &\multirow{2}{*}{$\left(\begin{array}{cccccc}    0.98 &0.22 \\-0.22 &0.98 \\ \end{array}\right)$}
          &\multirow{2}{*}{$\left(\begin{array}{cccccc}    3294 \\   3418  \end{array}\right)$}\\
$|J^P=7/2^+,3\rangle_{1D_{\rho }}$  &  &   &\\
\hline

$|J^P=1/2^-,1\rangle_{2P_\lambda}$  &(3371) &1    &3371\\
$|J^P=1/2^-,0\rangle_{2P_\lambda}$  &(3381) &1    &3381\\
$|J^P=1/2^-,1\rangle_{2P_\rho}$     &(3441) &1    &3441\\

$|J^P=1/2^-,1\rangle_{2P'_\lambda}$
                                &\multirow{2}{*}{$\left(\begin{array}{cccccc}   3523 &6   \\   6 & 	3530 \\ \end{array}\right)$}
                                &\multirow{2}{*}{$\left(\begin{array}{cccccc}   0.87 &-0.49  \\  0.49&  0.87 \\ \end{array}\right)$}
                                &\multirow{2}{*}{$\left(\begin{array}{cccccc}  3520   \\   3533 \\  \end{array}\right)$} \\
$|J^P=1/2^-,0\rangle_{2P'_\lambda}$  & &    &\\
$|J^P=1/2^-,1\rangle_{2P'_\rho}$     &(3575) & 1   &3575\\

$|J^P=3/2^-,2\rangle_{2P_\lambda}$
           &\multirow{3}{*}{$\left(\begin{array}{cccccc}  3393 &8 &4 \\ 8 &3400 &-26 \\ 4 &-26 &3463\\ \end{array}\right)$}
           &\multirow{3}{*}{$\left(\begin{array}{cccccc}    0.74 &0.63 &0.22  \\   -0.67 &0.70 &0.26 \\ 0.01 &-0.34 &0.94\\ \end{array}\right)$}
           &\multirow{3}{*}{$\left(\begin{array}{cccccc}    3401 \\   3383 \\ 3473\\ \end{array}\right)$}\\
$|J^P=3/2^-,1\rangle_{2P_\lambda}$  & &    &\\
$|J^P=3/2^-,1\rangle_{2P_\rho}$  & &    &\\

$|J^P=3/2^-,2\rangle_{2P'_\lambda}$
           &\multirow{2}{*}{$\left(\begin{array}{cccccc}  3522 &6  \\ 6 &3531 \\ \end{array}\right)$}
           &\multirow{2}{*}{$\left(\begin{array}{cccccc}    0.90 &-0.44  \\   0.44 &0.90 \\  \end{array}\right)$}
           &\multirow{2}{*}{$\left(\begin{array}{cccccc}    3518 \\   3534\\ \end{array}\right)$}\\
$|J^P=3/2^-,1\rangle_{2P'_\lambda}$  & &    &\\
$|J^P=3/2^-,1\rangle_{2P'_\rho}$  &(3589) & 1   &3589\\

$|J^P=5/2^-,2\rangle_{2P_\lambda}$  &(3412) &1    &3412\\
$|J^P=5/2^-,2\rangle_{2P'_\lambda}$  &(3528) &1    &3528\\

\hline\hline
\end{tabular}
\end{center}
\end{table*}

\begin{table*}
\begin{center}
\caption{ Partial strong decay widths~(MeV) of the excited $\Omega_c$ baryons. $\Gamma^{th}_{sum}$ stands for the sum of the partial decay widths.
The bold numbers are taken the experimental values.}
\label{Omegac decay widths}
\begin{tabular}{cccccccccccccc}\hline\hline
&State
&~~$\Gamma[\Xi_cK]$~
&~~$\Gamma[\Xi_c'K]$~
&~~$\Gamma[\Xi_c'^*K]$~
&~~$\Gamma[\Omega_c\eta]$~
&~~$\Gamma[\Omega_c(2770)\eta]$~
&~~$\Gamma[\Xi_c(2790)K]$~
&~~$\Gamma[\Xi_c(2815)K]$~
&~~$\Gamma_{sum}^{th}$~
&~~$\Gamma_{sum}^{exp}$~\\
\hline
& $\Omega_c (\mathbf{3000}) 1/2^-(1P)$                   &4.68   &$\cdot\cdot\cdot$      &$\cdot\cdot\cdot$      &$\cdot\cdot\cdot$        &$\cdot\cdot\cdot$       &$\cdot\cdot\cdot$       &$\cdot\cdot\cdot$   &4.68    &$3.83\pm 0.23^{+1.59}_{-0.29}$ \\
& $\Omega_c (\mathbf{3050}) 3/2^-(1P)$                   &1.00   &$\cdot\cdot\cdot$      &$\cdot\cdot\cdot$      &$\cdot\cdot\cdot$        &$\cdot\cdot\cdot$        &$\cdot\cdot\cdot$       &$\cdot\cdot\cdot$   &1.00    &$<1.8$\\
& $\Omega_c (\mathbf{3065}) 3/2^-(1P)$                   &2.77   &$\cdot\cdot\cdot$      &$\cdot\cdot\cdot$      &$\cdot\cdot\cdot$        &$\cdot\cdot\cdot$        &$\cdot\cdot\cdot$       &$\cdot\cdot\cdot$   &2.77    &$3.4^{+0.7}_{-0.8}$\\
& $\Omega_c (\mathbf{3120}) 3/2^-(1P)$                   &0.03   &0.06   &$\cdot\cdot\cdot$      &$\cdot\cdot\cdot$        &$\cdot\cdot\cdot$       &$\cdot\cdot\cdot$       &$\cdot\cdot\cdot$   &0.09    &$<2.5$\\
& $\Omega_c (\mathbf{3090}) 5/2^-(1P)$                            &7.31   &0.02   &$\cdot\cdot\cdot$      &$\cdot\cdot\cdot$        &$\cdot\cdot\cdot$        &$\cdot\cdot\cdot$       &$\cdot\cdot\cdot$   &7.33    &$8.48\pm 0.44^{+0.61}_{-1.62}$\\

& $\Omega_c (3041) 1/2^-$ $(1P)$            &84.35  &$\cdot\cdot\cdot$      &$\cdot\cdot\cdot$      &$\cdot\cdot\cdot$        &$\cdot\cdot\cdot$       &$\cdot\cdot\cdot$       &$\cdot\cdot\cdot$  &84.35   \\
& $\Omega_c (3090) 1/2^-$ $(1P)$            &0.19   &62.78   &$\cdot\cdot\cdot$      &$\cdot\cdot\cdot$       &$\cdot\cdot\cdot$       &$\cdot\cdot\cdot$       &$\cdot\cdot\cdot$  &62.97 \\
\hline
&$\Omega_c (3126) 1/2^+$ $(2S)$             &72.14  &24.84       &$\cdot\cdot\cdot$      &$\cdot\cdot\cdot$      &$\cdot\cdot\cdot$      &$\cdot\cdot\cdot$       &$\cdot\cdot\cdot$  &96.98\\
&$\Omega_c (3295) 1/2^+$ $(2S)$    &12.88  &16.54       &5.92   &0.10   &$\cdot\cdot\cdot$     &3.36     &$\cdot\cdot\cdot$  &38.80  \\
&$\Omega_c (3272) 1/2^+$ $(1D)$    &0.21   &5.38        &1.92   &0.03   &$\cdot\cdot\cdot$     &$\cdot\cdot\cdot$        &$\cdot\cdot\cdot$  &7.54\\
&$\Omega_c (3391) 1/2^+$ $(1D)$    &6.90   &1.19        &0.40   &0.02   &0.01  &10.53    &$\sim 0$ &19.05  \\
&$\Omega_c (3176) 3/2^+$ $(2S)$             &45.73  &8.03        &10.58  &$\cdot\cdot\cdot$         & $\cdot\cdot\cdot$     &$\cdot\cdot\cdot$    &$\cdot\cdot\cdot$  &64.34   &$50\pm 7^{+10}_{-20}$\\

&$\Omega_c (3317) 3/2^+$ $(2S)$             &14.33   &2.87      &17.29   &$\sim 0$   &$\sim 0$      &0.01   &0.83   &35.33    \\
&$\Omega_c (3282) 3/2^+$ $(1D)$             &$\sim 0$&1.84      &1.53   &0.04      &$\cdot\cdot\cdot$        &$\cdot\cdot\cdot$      & $\cdot\cdot\cdot$     & 3.41\\
&$\Omega_c (3276) 3/2^+$ $(1D)$             &0.17   &0.01       &4.93   &0.01  &$\cdot\cdot\cdot$        &$\cdot\cdot\cdot$     & $\cdot\cdot\cdot$    & 5.12\\
&$\Omega_c (3394) 3/2^+$ $(1D)$             &2.06   &0.69       &1.47   &0.02      &0.06    &0.09   & 5.22  & 9.61\\
&$\Omega_c (3422) 3/2^+$ $(1D)$             &3.21   &3.32       &1.75   &0.51      &0.01    &0.05 &15.88 &24.73\\

&$\Omega_c (3283) 5/2^+$ $(1D)$             &4.08   &1.63        &0.81   &$\sim 0$ &$\cdot\cdot\cdot$         &$\cdot\cdot\cdot$     &$\cdot\cdot\cdot$       &6.52\\
&$\Omega_c (3295) 5/2^+$ $(1D)$             &1.47   &0.08        &1.99   &$\sim 0$ &$\cdot\cdot\cdot$         &0.01  &$\cdot\cdot\cdot$       &3.54\\
&$\Omega_c (3379) 5/2^+$ $(1D)$             &8.51  &1.67        &1.26   &0.03     &0.01      &0.63  &0.13    &12.24\\
&$\Omega_c (3440) 5/2^+$ $(1D)$             &0.07   &2.52        &3.58  &0.08     &0.47      &0.06  &0.44    & 7.22 \\

&$\Omega_c (3294) 7/2^+$ $(1D)$             &5.42  &0.49        &0.43   &$\sim 0$  &0.05         &$\sim 0$ &$\cdot\cdot\cdot$    &6.39\\
&$\Omega_c (3418) 7/2^+$ $(1D)$             &12.40  &1.84        &2.80   &$\cdot\cdot\cdot$     &0.03      &0.04     &1.02 &18.13\\
\hline
&$\Omega_c (3371) 1/2^-$ $(2P)$             &$\sim 0$ &1.31      &1.03    &34.18     & 0.09     &1.10      &14.66     &52.37  \\
&$\Omega_c (3381) 1/2^-$ $(2P)$             &2.89    &$\sim 0$  &$\sim 0$&$\sim 0$ &$\sim 0$  &0.03     &0.48      &3.40 \\
&$\Omega_c (3441) 1/2^-$ $(2P)$             &$\sim 0$ &1.82      &6.49   &37.77     &0.58      &$\sim 0$ &$\sim 0$  &46.66 \\
&$\Omega_c (3520) 1/2^-$ $(2P)$    &0.94     &2.02      &4.59    &1.53     &0.53      &6.21     &15.52     &31.34  \\
&$\Omega_c (3533) 1/2^-$ $(2P)$    &3.19     &0.68      &1.69    &0.61     &0.19     &18.23    &3.03      &27.62\\
&$\Omega_c (3575) 1/2^-$ $(2P)$             &$\sim 0$ &1.82      &6.49    &26.47     &5.17     & $\sim 0$& $\sim 0$ &39.95\\
&$\Omega_c (3401) 3/2^-$ $(2P)$    &$\sim 0$ &0.02    &$\sim 0$   &0.01   &4.40      &4.38    &0.20      &9.01   \\
&$\Omega_c (3383) 3/2^-$ $(2P)$    &0.02     &0.69     &1.10    &0.49     &4.03      &0.96     &2.94    &10.23\\
&$\Omega_c (3473) 3/2^-$ $(2P)$    &$\sim 0$ &0.37     &4.50   &1.11    &7.78      &0.04     &0.60      &14.40\\
&$\Omega_c (3518) 3/2^-$ $(2P)$    &1.93     &3.82      &1.30    &2.94     &0.12      &6.15     &11.52    &27.78\\
&$\Omega_c (3534) 3/2^-$ $(2P)$    &0.40     &0.13      &5.11    &0.12     &4.19      &12.53    &9.79     &32.27  \\
&$\Omega_c (3589) 3/2^-$ $(2P)$             &$\sim 0$ &1.25      &1.20    &4.37     &26.91      &$\sim 0$ &$\sim 0$  &33.73        \\

&$\Omega_c (3412) 5/2^-$ $(2P)$             &1.12     &0.06      &0.05    &0.12     &0.28      &0.07     &1.18    &2.88\\
&$\Omega_c (3528) 5/2^-$ $(2P)$             &1.86     &0.98      &3.60    &0.38     &1.08      &0.02    &16.32     &24.24  \\

\hline\hline
\end{tabular}
\end{center}
\end{table*}

\subsection{$1S$-wave states}

According to the quark model classification, there are two $1S$-wave states
$|J^P=1/2^+,1\rangle_{1S}$ and $|J^P=3/2^+,1\rangle_{1S}$.
They should correspond to the $\Omega_c(2695)1/2^+$ and $\Omega_c(2770)3/2^+$
first observed in the early WA62 and Belle experiments~\cite{Biagi:1984mu,Solovieva:2008fw}, respectively.
Their measured masses can be well described in the quark potential model.

\subsection{$1P$-wave states}

In the $1P$-wave states, there are five $\lambda$-mode excitations and two $\rho$-mode excitations
according to the quark model classification. From Table~\ref{Omegac mass spectrum comparison}, it is seen that
there is slight mixing between different configurations with the same spin-parity numbers.
The masses for the $1P$ states scatter
in the range of $\sim3.0-3.1$ GeV. In the $1P$ states with $J^P=1/2^-$ and $3/2^-$,
the two high-lying states are dominated by the $\rho$-mode excitations, whose masses are about
$50-80$ MeV larger than those of the $\lambda$-mode dominant states.

From the point of view of the mass, the five narrow resonances, $\Omega_c(3000)$, $\Omega_c(3050)$, $\Omega_c(3065)$, $\Omega_c(3090)$, and $\Omega_c(3120)$, observed in experiments are good candidates of the $1P$-wave $\Omega_c$ excitations.
For the convenience of comparing with our theoretical results, the PDG averaged masses and widths of these resonances are presented as follows~\cite{ParticleDataGroup:2022pth},
\begin{eqnarray}
m[\Omega_c(3000)]&=&3000.46\pm0.25~\mathrm{MeV},\nonumber\\
\Gamma[\Omega_c(3000)]&=&3.83\pm 0.23^{+1.59}_{-0.29}~\mathrm{MeV},\nonumber\\
m[\Omega_c(3050)]&=&3050.17\pm0.19~\mathrm{MeV},\nonumber\\
\Gamma[\Omega_c(3050)]&<&1.8~\mathrm{MeV},\nonumber\\
m[\Omega_c(3065)]&=&3065.58\pm0.21~\mathrm{MeV},\nonumber\\
\Gamma[\Omega_c(3065)]&=&3.4^{+0.7}_{-0.8}~\mathrm{MeV},\nonumber\\
m[\Omega_c(3090)]&=&3090.15\pm0.26~\mathrm{MeV},\nonumber\\
\Gamma[\Omega_c(3090)]&=&8.48\pm 0.44^{+0.61}_{-1.62}~\mathrm{MeV},\nonumber\\
m[\Omega_c(3120)]&=&3118.98\pm0.35^{+0.09}_{-0.23}~\mathrm{MeV},\nonumber\\
\Gamma[\Omega_c(3120)]&<&2.5~\mathrm{MeV}.\nonumber
\end{eqnarray}
By combining the mass spectrum with the decay properties given in Table~\ref{Omegac decay widths},
we further give some discussions about the nature of these newly observed resonances as follows.

\subsubsection{$\Omega_c(3000)$ as a $1P$ state with $J^P=1/2^-$}


There are three $1P$ states with $J^P=1/2^-$. In these states there is slight configuration mixing between $|J^P=1/2^-,1\rangle_{1P_\lambda}$, $|J^P=1/2^-,0\rangle_{1P_\lambda}$, and $|J^P=1/2^-,1\rangle_{1P_\rho}$, the details can be seen in Table~\ref{Omegac mass spectrum comparison}.

The $\Omega_c(3000)$ resonance favors the assignment of the lowest $J^P=1/2^-$ state, whose dominant component is $|J^P=1/2^-,1\rangle_{1P_\lambda}$ ($92\%$).
With this assignment, both the theoretical mass and width,
\begin{equation}
M\simeq3025~ \mathrm{MeV}, ~\Gamma \simeq 4.68~\mathrm{MeV},
\end{equation}
are consistent with the observations of $\Omega_c(3000)$. It is emphasized that
the slight configuration mixing for the lowest $J^P=1/2^-$ state is crucial for understanding its strong decays.
The $\Xi_cK$ channel, as the only OZI-allowed two body strong decay channel,
is forbidden if the small $|J^P=1/2^-,0\rangle_{1P_\lambda}$ component ($\sim6\%$) is absent.
The assignment (dominated by the $1^2P_\lambda$ component in the $L$-$S$ coupling scheme) in the present work is consistent with that of our previous work~\cite{Wang:2017hej} and other suggestions in Refs.~\cite{Ortiz-Pacheco:2023kjn,Yu:2022ymb,Garcia-Tecocoatzi:2022zrf,Santopinto:2018ljf,Agaev:2017lip,Chen:2017gnu,Wang:2017zjw,Weng:2024roa}.
The $J^P=1/2^-$ assignment for the $\Omega_c(3000)$ is also supported by the recent LHCb measurements~\cite{LHCb:2021ptx}.

For the $J^P=1/2^-$ state dominated by
the $|J^P=1/2^-,0\rangle_{1P_\lambda}$ configuration, the mass and width
are predicted to be
\begin{equation}
M\simeq3041~ \mathrm{MeV}, ~\Gamma \simeq 84~\mathrm{MeV}.
\end{equation}
The decay width is nearly saturated by the $\Xi_cK$ channel. This broad state predicted in the present work may lie about 20 MeV
above the $\Omega_c(3000)$ resonance.
It should be mentioned that the $|J^P=1/2^-,0\rangle_{1P_\lambda}$ dominant state may be shifted below the mass threshold of $\Xi_cK$ due to the coupled-channel effects~\cite{Luo:2021dvj,Zhang:2025gar}.
Thus, it may be a very narrow state since the OZI-allowed $\Xi_cK$ channel is forbidden.
However, based on the framework of the unquenched quark model~\cite{Zhong:2022cjx,Ni:2023lvx}, we estimate the mass shift caused
by $\Xi_cK$ hadronic loop, it is found the coupled-channel effects on the mass shift is no more than $10$ MeV.

The highest $J^P=1/2^-$ state is a broad state dominated by the
$|J^P=1/2^-,1\rangle_{1P_\rho}$ configuration. The mass and width
are predicted to be
\begin{equation}
M\simeq3090~ \mathrm{MeV}, ~\Gamma \simeq 63~\mathrm{MeV}.
\end{equation}
The strong decays are governed by the $\Xi_c'K$ channel. There is a small decay rate ($\sim0.3\%$) into
the $\Xi_cK$ channel due to small mixing with the $\lambda$-mode configurations as shown in Table~\ref{Omegac mass spectrum comparison}.
Our predicted mass is in good agreement with that in a recent work~\cite{Weng:2024roa}.
It should be mentioned that the narrow resonance $\Omega_c(3090)$ observed in experiments
cannot be assigned as this $\rho$-mode dominant $J^P=1/2^-$ state although the measured
mass are close to the theoretical value, because the observed decay width and decay mode
are inconsistent with the theoretical predictions.

As a whole, the observations of $\Omega_c(3000)$ can be well explained with
the $|J^P=1/2^-,1\rangle_{1P_\lambda}$ dominant state. The slight mixing with
the $|J^P=1/2^-,0\rangle_{1P_\lambda}$ is crucial for understanding its narrow width.
To further test the nature of $\Omega_c(3000)$ and various model predictions,
we expect to look for the other two missing $J^P=1/2^-$ states in the $\Xi_cK$ and
$\Xi_c'K$ final states around energy region of $\sim3.0-3.1$ GeV in future experiments.

\subsubsection{$\Omega_c(3050,3065,3120)$ as $1P$ states with $J^P=3/2^-$}

There are three $1P$ states with $J^P=3/2^-$. In these states there is significant configuration mixing
between $|J^P=3/2^-,1\rangle_{1P_\lambda}$, $|J^P=3/2^-,2\rangle_{1P_\lambda}$, and $|J^P=3/2^-,1\rangle_{1P_\rho}$,
the details can be seen in Table~\ref{Omegac mass spectrum comparison}.

The narrow resonance $\Omega_c(3050)$ observed in the $\Xi_c^+K^-$ channel favors the lowest $J^P=3/2^-$ state, whose dominant component is $|J^P=3/2^-,1\rangle_{1P_\lambda}$ ($\sim52\%$).
This state also has significant components of $|J^P=3/2^-,2\rangle_{1P_\lambda}$ ($\sim35\%$) and $|J^P=3/2^-,1\rangle_{1P_\rho}$ ($\sim13\%$).
The $\Xi_cK$ is the only OZI-allowed two body strong decay channel.
With this assignment, both the theoretical mass and width,
\begin{equation}
M\simeq3045~ \mathrm{MeV}, ~\Gamma \simeq 1.0~\mathrm{MeV},
\end{equation}
are consistent with the data $M_{exp}\simeq 3050$ MeV and $\Gamma_{exp}< 1.8$ MeV.
Furthermore, the $J^P=3/2^-$ assignment for the $\Omega_c(3050)$ is
supported by the subsequent LHCb measurements~\cite{LHCb:2021ptx}.
Our assignment is also consistent with the suggestions
in Refs.~\cite{Chen:2017gnu,Wang:2017hej,Weng:2024roa,Yu:2023bxn,Yu:2022ymb},
although there are many other explanations in the literature (see the review work~\cite{Cheng:2021qpd}).
It should be pointed out that the configuration mixing is crucial
for understanding the strong decay properties of $\Omega_c(3050)$ observed in experiments.
If it is a pure $|J^P=3/2^-,1\rangle_{1P_\lambda}$ state, the strong decay channel $\Xi_cK$ is forbidden in the heavy quark symmetry limit.

The narrow resonance $\Omega_c(3065)$ observed in the $\Xi_c^+K^-$ channel favors the moderate mass state, whose dominant component is $|J^P=3/2^-,2\rangle_{1P_\lambda}$ ($\sim65\%$).
This state has a significant component of $|J^P=3/2^-,1\rangle_{1P_\lambda}$ ($\sim30\%$).
The $\Xi_cK$ is the only OZI-allowed two body strong decay channel.
With this assignment, both the theoretical mass and width,
\begin{equation}
M\simeq3064~ \mathrm{MeV}, ~\Gamma \simeq 2.8~\mathrm{MeV},
\end{equation}
are consistent with the data $M_{exp}\simeq 3066$ MeV and $\Gamma_{exp}\simeq3.4^{+0.7}_{-0.8}$ MeV.
Furthermore, the $J^P=3/2^-$ assignment for the $\Omega_c(3065)$ is
supported by the subsequent LHCb measurements~\cite{LHCb:2021ptx}. Our assignment is also consistent with the suggestions
in Refs.~\cite{Chen:2017gnu,Wang:2017hej,Weng:2024roa,Yu:2023bxn,Yu:2022ymb}. It should be mentioned
that the configuration mixing is also crucial for understanding the mass splitting between
$\Omega_c(3050)$ and $\Omega_c(3065)$. If they are pure states in the $j$-$j$ coupling scheme,
the mass order predicted in theory should be reversed as shown in Table~\ref{Omegac mass spectrum comparison}.

The narrow resonance $\Omega_c(3120)$ observed in the $\Xi_c^+K^-$ channel favors the highest state, whose dominant component is $|J^P=3/2^-,1\rangle_{1P_\rho}$ ($\sim82\%$).
This state has a significant component of $|J^P=3/2^-,1\rangle_{1P_\lambda}$ ($\sim18\%$),
and dominantly decays into $\Xi_cK$ and $\Xi_c'K$ channels with a ratio of $\Gamma[\Xi_cK]:\Gamma[\Xi_c'K]\simeq 1:2$.
With this assignment, both the theoretical mass and width,
\begin{equation}
M\simeq3126~ \mathrm{MeV}, ~\Gamma \simeq 0.09~\mathrm{MeV},
\end{equation}
are consistent with the data $M_{exp}\simeq 3119$ MeV and $\Gamma_{exp}<2.5$ MeV.
The $\Omega_c(3120)$ as a $J^P=3/2^-$ state dominated by
the $\rho$-mode excitation is also suggested in a recent work~\cite{Weng:2024roa}
based on a mass spectrum analysis.
It should be pointed out that the configuration mixing is crucial
for understanding the strong decay properties of $\Omega_c(3120)$ observed in experiments.
If it is a pure $\rho$-mode excitation $|J^P=3/2^-,1\rangle_{1P_\rho}$, the strong decay channel $\Xi_cK$ is forbidden.

As a whole, the $\Omega_c(3050)$ and $\Omega_c(3065)$ resonances can be assigned
as the $\lambda$-mode excitation dominant states with $J^P=3/2^-$.
The $\Omega_c(3120)$ favors the $\rho$-mode excitation dominant state with $J^P=3/2^-$.
The configuration mixing is crucial for understanding their strong decay properties and mass splitting.
It should be emphasized that for the $\Omega_c(3050,3065,3120)$, there are many other explanations in the literature (see the review work~\cite{Cheng:2021qpd}), more studies are needed in both theory and experiments.

\subsubsection{$\Omega_c(3090)$ as the $1P$ state with $J^P=5/2^-$}

In the $1P$ states, there is only one $J^P=5/2^-$ state $|J^P=5/2^-,2\rangle_{1P_\lambda}$,
which is a $\lambda$-mode excitation. It is a narrow width state dominantly decaying
into the $\Xi_cK$ channel.

The $\Omega_c(3090)$ observed in experiments should correspond to the $1P$ state with $J^P=5/2^-$.
With this assignment, both the mass and width predicted in theory,
\begin{equation}
M\simeq3082~ \mathrm{MeV}, ~\Gamma \simeq 7.3~\mathrm{MeV},
\end{equation}
are consistent with the observations $M_{exp}\simeq 3090$ MeV and $\Gamma_{exp}\simeq 8.5$ MeV.
Our assignment for the $\Omega_c(3090)$ is consistent with
the suggestions in Refs.~\cite{Wang:2017hej,Weng:2024roa,Li:2024zze,Zhao:2017fov,Yu:2022ymb,Yu:2023bxn}
and the recent LHCb observations~\cite{LHCb:2021ptx}.

\subsection{$1D$- and $2S$-wave states}

According to the quark model classification, there are four $2S$-, six $1D_{\lambda}$-,
and six $1D_{\rho}$-wave configurations. The mixing between different
configurations with the same spin-parity numbers is considered.
Our results are given in Table~\ref{Omegac mass spectrum comparison}.
It is found that there is significant configuration mixing in some states.
For example, the $|J^P=3/2^+,1\rangle_{2S }$ dominant state $\Omega_c(3295)1/2^+$
containing significant components of $|J^P=3/2^+,1\rangle_{1D_{\lambda}}$ ($\sim27\%$).

From Table~\ref{Omegac mass spectrum comparison} and Fig~\ref{omegac mass spectrum},
one can see that the two high-lying $2S$-wave dominant states have a mass of $\sim 3.31$ GeV,
while the other two low-lying $2S$ excitations have a mass of $\sim3.15$ GeV.
There is a gap of $\sim200$ MeV between the low-lying and high-lying $2S$ states.
It should be emphasized that the two low-lying $2S$ states are dominated by the
$\lambda-$mode radial excitation, while the two high-lying states are dominated by the
$\rho-$mode radial excitation. For the $1D$ excitations, the masses of the $\lambda$-mode dominant states
are $\sim3.3$~GeV,
while the masses of the $\rho$-mode dominant states are $\sim3.4$~GeV.
The mass of the $\rho$-mode excitation is about 100~MeV heavier than the $\lambda$-mode excitation.

Furthermore, from Table~\ref{Omegac decay widths}, one can obtain the decay properties of $2S$ and $1D$ excitations.
For the $2S$-wave dominant states, the decay widths are relatively broad, $\Gamma\sim30-100$ MeV. They mainly decay into the
$\Xi_cK$, $\Xi_c'K$, and/or $\Xi_c'^*K$ channels. Their partial widths often reach up to $\mathcal{O}(10)$~MeV.
For the $1D$-wave $\lambda$-mode dominant states, the decay widths may be relatively narrow,
the sum of the partial widths of the $\Xi_cK$, $\Xi_c'K$, and $\Xi_c'^*K$ channels is only several MeV.
In the $\lambda$-mode dominant states, only the $\Omega_c(3283)5/2^+$ and $\Omega_c(3294)7/2^+$ have a relatively large partial width ($\sim 4$ MeV)
into the $\Xi_cK$ channel. For the $1D$-wave $\rho$-mode dominant states, it is found that the $\Omega_c(3391)1/2^+$, $\Omega_c(3422)3/2^+$, $\Omega_c(3418)7/2^+$ may have a relatively broad width, $\Gamma\sim20$ MeV.
The decay widths for the other $\rho$-mode dominant states may be relatively narrow,
the sum of the partial widths of the $\Xi_cK$ and $\Xi_c'K$, and $\Xi_c'^*K$ channels
is $\sim10$ MeV.

For the $1D$- and $2S$-wave states, several candidates, such as $\Omega_c(3185)$ and $\Omega_c(3327)$, have been observed in recent LHCb experiments~\cite{LHCb:2017uwr,LHCb:2023sxp}. Based on the obtained decay properties and mass spectrum,
some discussions about these newly observed states are given as follows.

\subsubsection{$\Omega_c(3185)$ as a radial excitation with $J^P=3/2^+$}

The $\Omega_c(3185)$ was observed in the $\Xi^+_cK^-$ channel by the LHCb Collaboration in 2023~\cite{LHCb:2023sxp}.
The measured mass and width are~\cite{ParticleDataGroup:2022pth}
\begin{eqnarray}
m[\Omega_c(3185)]&=&3185\pm 1.7^{+7.4}_{-0.9}~\mathrm{MeV},\nonumber\\
\Gamma[\Omega_c(3185)]&=&50\pm 7^{+10}_{-20}~\mathrm{MeV}.\nonumber
\end{eqnarray}
It is found that the $\lambda$-mode radial excitation $|J^P=3/2^+,1\rangle_{2S}$
favors the assignment of $\Omega_c(3185)$. The predicted mass and width,
\begin{equation}
M\simeq3176~ \mathrm{MeV}, ~\Gamma \simeq 64~\mathrm{MeV},
\end{equation}
are in good agreement with the observations. This assignment for the $\Omega_c(3185)$ is consistent with
the that suggested in the literature~\cite{Yu:2023bxn,Weng:2024roa,Jakhad:2023mni}.
The $|J^P=3/2^+,1\rangle_{2S}$ dominantly decays into the $\Xi_cK$, $\Xi_c'K$,
and $\Xi_c'^*K$ channels with partial width ratios
\begin{equation}
\Gamma[\Xi_cK]:\Gamma[\Xi_c'K]:\Gamma[\Xi_c'^*K]=1.0:0.17:0.23.
\end{equation}
The $\Omega_c(3185)$ resonance, as the $|J^P=3/2^+,1\rangle_{2S}$ assignment,
may be most likely to be observed in the $\Xi_c'K$ and $\Xi_c'^*K$ channels.

If the $\Omega_c(3185)$ corresponds to $|J^P=3/2^+,1\rangle_{2S}$ indeed, its spin partner
$|J^P=1/2^+,1\rangle_{2S}$ is most likely
to be observed in future experiments. The mass and width of this state are
predicted to be
\begin{equation}
M\simeq3126~ \mathrm{MeV}, ~\Gamma \simeq 97~\mathrm{MeV}.
\end{equation}
This broad state mainly decays into the $\Xi_cK$ and $\Xi_c'K$ channels with
branching fractions $\sim74\%$ and $\sim26\%$, respectively.

Finally, it should be mentioned that in previous work~\cite{Wang:2017hej},
very narrow widths ($\Gamma\sim 1$ MeV) were predicted for the two low-lying $2S$
states with $J^P=1/2^+$ and $3/2^+$ based on the simple harmonic oscillator wave functions.
Thus, the $\Omega_c(3120)$ was suggested as the $2S$ assignment with $J^P=1/2^+$ or $3/2^+$.
However, in the present work we obtain broad widths for these two low-lying $2S$
states. The main reasons are: (i) we adopt the genuine wave functions obtained from
the quark potential model, in which besides the dominant $\lambda$-mode component
there is also a significant $\rho$-mode component; (ii) we include an additional relativistic correction term
in the strong decay operators, which is found to play an important role in the strong
decays of the radially excited hadrons~\cite{Arifi:2021orx,Arifi:2022ntc,Ni:2023lvx,Zhong:2024mnt}.

\subsubsection{$\Omega_c(3327)$ as a $1D$ state with $J^P=5/2^+$ or $J^P=7/2^+$}

The $\Omega_c(3327)$ also was observed in the $\Xi^+_cK^-$ channel by the LHCb Collaboration in 2023~\cite{LHCb:2023sxp}.
The measured mass and width of this state are~\cite{ParticleDataGroup:2022pth},
\begin{eqnarray}
m[\Omega_c(3327)]&=&3327.1\pm 1.2^{+0.1}_{-1.3}~\mathrm{MeV},\nonumber\\
\Gamma[\Omega_c(3327)]&=&20\pm 5^{+13}_{-1}~\mathrm{MeV}.\nonumber
\end{eqnarray}
Combining the mass and the main decay channel of $\Xi^+_cK^-$, it is found that
the two high-lying states dominated by the $2S$-wave excitations~($\Omega_c(3295)1/2^+$ and $\Omega_c(3317)3/2^+$),
and three states dominated by the $1D$-wave $\lambda$-mode excitations~($\Omega_c(3283)5/2^+$,
$\Omega_c(3295)5/2^+$, and $\Omega_c(3294)7/2^+$)
may be candidates of $\Omega_c(3327)$.

It should be noted that these candidates with a mass of $\sim3.3$ GeV also possibly decay into
the $\Xi D$ and $\Xi D^*$ channels. Using the quark pair creation model~\cite{Micu:1968mk,LeYaouanc:1972vsx,LeYaouanc:1973ldf},
Li-Ye Xiao has estimated the partial widths of the $\Xi D$ and $\Xi D^*$ channels~\cite{Xiao:2025}.
For the two $2S$ dominant states $\Omega_c(3295)1/2^+$ and $\Omega_c(3317)3/2^+$,
it is found that the widths may be too broad. The total widths are predicted to be $\sim100$ MeV,
which is not in good agreement with the narrow width nature of $\Omega_c(3327)$.

Then, we focus on the three $1D$-wave dominant candidates, $\Omega_c(3283)5/2^+$,
$\Omega_c(3295)5/2^+$, and $\Omega_c(3294)7/2^+$.
It is interesting to find that the $\Omega_c(3327)$ seems to favor the $J^P=7/2^+$ state
$\Omega_c(3294)7/2^+$, although the $J^P=5/2^+$ states $\Omega_c(3283)5/2^+$ and
$\Omega_c(3295)5/2^+$ cannot be excluded. As shown in Table~\ref{Omegac decay widths 2}, the branching fraction of
$Br[\Omega_c(3294)7/2^+\to \Xi D]\simeq 25\%$ is relatively large.
To confirm the nature of $\Omega(3327)$, the $\Xi D$ invariant mass spectrum around 3.3 GeV
is expected to be observed in future experiments.
Finally, it should be pointed out that the $\Omega_c(3327)$ is also suggested to
be a candidate of the $1D$-wave excitations in the literature~\cite{Li:2024zze,Luo:2023sra,Yu:2023bxn,Wang:2023wii,Jakhad:2023mni,Pan:2023hwt}, however, the assignments of the spin-parity numbers are inconsistent with each other.

\begin{table*}
\begin{center}
\caption{The strong decay widths (MeV) of $\Omega_c(3295)1/2^+(2S)$, $\Omega_c(3317)3/2^+(2S)$, $\Omega_c(3283)5/2^+(1D)$,
$\Omega_c(3295)5/2^+(1D)$, and $\Omega_c(3294)7/2^+(1D)$ as the candidates of $\Omega_c(3327)$. For a comparison with the observations,
the masses of the initial $\Omega_c$ excitations are taken as the experimental value, 3327 MeV.}
\label{Omegac decay widths 2}
\begin{tabular}{cccccccccccccc} \hline \hline
& States &  $\Gamma[\Xi_c K]$ &  $\Gamma[\Xi_c' K]$  & $\Gamma[\Xi_c'^* K]$
& $\Gamma[\Omega_c  \eta]$    & $\Gamma[\Omega_c(2770)  \eta]$
& $\Gamma[\Omega_c(2790)  K]$    & $\Gamma[\Omega_c(2815)  K]$
& $\Gamma[\Xi D]$& $\Gamma[\Xi D^*]$

& ~~$\Gamma^{total}$~~&$\Gamma^{exp}$~\cite{ParticleDataGroup:2022pth} \\
\hline
& $\Omega_C(3295)1/2^+(2S)$  & 12.7 &18.3 &7.2  &0.2      &$\sim 0$ &5.4      &$\sim 0$ &60.2     &1.7      & 105.7  & $20^{+14}_{-5}$\\
& $\Omega_C(3317)3/2^+(2S)$  & 14.2 &3.0  &18.2 &$\sim 0$ &$\sim 0$ &$\sim 0$ &1.2      &55.5     &2.3      & 94.4   &\\
& $\Omega_C(3283)5/2^+(1D)$  & 6.1  &3.1  &1.0  &$\sim 0$ &$\sim 0$ &0.2      &$\sim 0$ &0.7      &0.2      & 11.3   &\\
& $\Omega_C(3295)5/2^+(1D)$  & 2.0  &0.1  &2.5  &$\sim 0$ &$\sim 0$  &0.3     &$\sim 0$ &0.5      &0.7      & 6.1    &\\
& $\Omega_C(3294)7/2^+(1D)$  & 7.3  &0.8  &0.8  &$\sim 0$ &$\sim 0$ &$\sim 0$ &0.1      &3.2      &$\sim 0$ & 12.2    &\\
\hline \hline
\end{tabular}
\end{center}
\end{table*}

\subsection{$2P$-wave states}

According to the quark model classification, there are fourteen $2P$-wave configurations.
The mixing between different configurations is also considered. The results are given in Table~\ref{Omegac mass spectrum comparison}.
It is found that the configuration mixing between the $\lambda$- and $\rho$-mode orbital excitations
is negligibly small. The masses of the $2P_\lambda$ and $2P_\rho$ states are in
the ranges of $\sim3.3-3.5$~GeV and $\sim3.4-3.6$~GeV, respectively.
Furthermore, we find that the states dominated by the $\rho$-mode radial excitation (denoted by $2P'$ with subscripts) are about 120-170~MeV
heavier than the corresponding $2P$ states dominated by the $\lambda$-mode radial excitation.

From Table~\ref{Omegac decay widths}, we can obtain the decay properties of $2P$ excitations.
In these fourteen states, only three states $\Omega_c(3381)1/2^-$, $\Omega_c(3533)1/2^-$, $\Omega_c(3518)3/2^-$,
$\Omega_c(3412)5/2^-$, and $\Omega_c(3528)5/2^-$ have significant decay rates into the $\Xi_cK$ channel.
It is interesting to find that several states, such as $\Omega_c(3371)1/2^-$,
$\Omega_c(3441)1/2^-$, and $\Omega_c(3575)1/2^-$, have large decay rates into
the $\Omega_c\eta$. By observing the invariant mass spectrum
of $\Omega_c\eta$ in the mass region of $\sim3.3-3.6$ GeV, one may find some signals of the $2P$ states.

\section{summary} \label{summary}

In this work, we carry out a unified study of the mass spectrum and strong decays
of the $\Omega_c$ baryon states up to the $2P$ excitations within a
constituent quark model. In the mass spectrum calculations,
we consider the dynamical mixing between different configurations with the same spin-parity numbers.
While in the strong decay evaluations, we incorporated the contribution of the
relativistic correction term as an improvement. It is found that the widths, and masses together with the hyperfine mass splittings, for the $\Omega_c(X)$
resonances observed in experiments can be well described within the quark model.
The configuration mixing is crucial for understanding the strong decay properties and mass splittings,
while the the relativistic correction term of the strong transition operator plays an important role in the states dominated by the radial excitations. Some key results from this study are emphasized as follows.

\begin{itemize}
\item
The narrow resonances $\Omega_c(3000)$, $\Omega_c(3050)$, $\Omega_c(3065)$, and $\Omega_c(3090)$
favor the $1P$-wave $\lambda$-mode dominant states with $J^P=1/2^-$, $3/2^-$, $3/2^-$, and $5/2^-$, respectively.
While the narrow resonance $\Omega_c(3120)$ favors the $1P$-wave $\rho$-mode dominant state with $J^P=3/2^-$ rather than a radial excitation suggested in our previous work when the relativistic correction term of the strong transition operator is included.

\item In the $1P$-wave states, there are two missing broad $J^P=1/2^-$ resonances lying $\sim20$ and $\sim 70$ MeV above $\Omega_c(3000)$. They may have potentials to be observed in the $\Xi_cK$ and $\Xi_c'K$ channels, respectively.

\item The newly observed state $\Omega_c(3185)$ can be assigned as the $2S$ state with $J^P=3/2^+$, $\Omega_c|J^P=3/2^+,1\rangle_{2S}$,
which is dominated by the $\lambda$-mode radial excitation. Its spin partner
$\Omega_c|J^P=1/2^+,1\rangle_{2S}$(3126) is likely
to be observed in the $\Xi_cK$ channel as well.

\item
The newly observed state $\Omega_c(3327)$ seems to favor the $1D$ state
with $7/2^+$ dominated by the $\lambda$-mode excitation, although the other two $1D$-wave $J^P=5/2^+$ states $\Omega_c(3283)5/2^+$ and
$\Omega_c(3295)5/2^+$ cannot be excluded. To determine nature of $\Omega(3327)$, the $\Xi D$ channel is worth observing in future experiments.

\item In the mass region
of $\sim3.3-3.6$ GeV, some $2P$ states are most likely to be observed in the invariant mass spectrum
of $\Omega_c\eta$.
\end{itemize}

\section*{Acknowledgement}

The authors would like to thank Ru-Hui Ni for very helpful discussions of the strong decay.
This work is supported by the National Natural Science Foundation of China (Grants No.12105203, No.12175065, No.12235018, No.12205216, No.12205026, and No.12005013).

\bibliographystyle{unsrt}

\end{document}